\documentclass[fleqn,usenatbib]{rasti}

\usepackage{newtxtext,newtxmath}

\usepackage[T1]{fontenc}

\DeclareRobustCommand{\VAN}[3]{#2}
\let\VANthebibliography\thebibliography
\def\thebibliography{\DeclareRobustCommand{\VAN}[3]{##3}\VANthebibliography}

\usepackage{graphicx}	
\usepackage{amsmath}	

\usepackage{caption}

\newcommand{\figp}[1]{~\hyperref[#1]{\mbox{(Fig.~\ref*{#1})}}}
\newcommand{\fig}[1]{~\hyperref[#1]{\mbox{Fig.~\ref*{#1}}}}
\newcommand{\Figp}[2]{~\hyperref[#1]{\mbox{(#2~\ref*{#1})}}}
\newcommand{\Fig}[2]{~\hyperref[#1]{\mbox{#2~\ref*{#1}}}}

\newcommand{\vect}[1]{\boldsymbol{#1}} 
\newcommand{\mat}[1]{\boldsymbol{\mathsf{#1}}} 
\newcommand{\um}{\ensuremath{\text{\textmu m}}} 
\definecolor{ForestGreen}{RGB}{48, 163, 7}

\providecommand{\lvlink}[3][]{\csname @bsphack\endcsname\csname @esphack\endcsname}

\DeclareRobustCommand{\stickynote}[1]{\futurelet\lvSNnext\lvSNcheck}
\providecommand{\lvSNcheck}{\ifx\lvSNnext\bgroup\expandafter\lvSNtwo\else\expandafter\lvSNone\fi}
\providecommand{\lvSNtwo}[1]{\lvSNone}
\providecommand{\lvSNone}{\csname @bsphack\endcsname\csname @esphack\endcsname}

\title[Gaussian perturbation interpretability]{{Gaussian perturbation interpretability for regression techniques} to understand retrieval black-boxes}

\author[J. D. Clarke et al.]{
Jools D. Clarke,$^{1}$\thanks{E-mail: j.d.clarke@ucl.ac.uk}
Kai Hou Yip$^{2}$
and Nikolaos Nikolaou$^{1}$
\\
$^{1}$Department of Physics and Astronomy, University College London, Gower Street, WC1E 6BT London, United Kingdom\\
$^{2}$Department of Physics, King’s College London, Strand Building, Strand, WC2R 2LS London, United Kingdom
}

\date{Accepted XXX. Received YYY; in original form ZZZ}

\pubyear{\the\year{}}

\begin{document}
\label{firstpage}
\pagerange{\pageref{firstpage}--\pageref{lastpage}}
\maketitle

\begin{abstract}
Characterizing the atmospheres of large numbers of exoplanets from their observed spectra can answer key questions about planetary populations. Traditional retrievals are resource-intensive; machine-learning methods offer a fast alternative, but their complexity renders them black-boxes. Understanding how they reach their predictions allows us to inspect them for biases and verify predictions are based on physically plausible spectral features. Commonly used perturbation-based methods for explaining machine learning struggle with the strong correlations between inputs in spectra. We introduce Gaussian perturbation methodology for explaining black-box regression models with one-dimensional structured inputs, which addresses these issues by leveraging physical knowledge of spectral correlation. We compare it with naive perturbation of one wavelength bin at a time. Although traditional retrievals are built on physical forward models, their mapping from spectrum to retrieved parameters is a black-box. We therefore first demonstrate this methodology using \textsc{TauREx-III} as a proxy black-box retrieval on \textit{HST} observations of HD~209458~b and \textit{JWST} observations of \mbox{HAT-P-26}~b, showing the efficacy of Gaussian perturbation-based interpretability against a well-established, physically motivated retrieval framework, before applying it to a convolutional neural network retrieval of \mbox{HAT-P-26}~b. For the gases the retrievals constrain, the importance from Gaussian perturbation agrees with the expected importance, including at the core of the $\mathrm{CO_2}$ band of \mbox{HAT-P-26}~b, where a degeneracy inverts the response. The method shows which parts of a spectrum a retrieval relies on, which the physics of its forward model alone does not, and shows the neural network has learned the same degeneracy.

\end{abstract}


\section{Introduction}
\label{sec:intro}
Since the first transiting exoplanet observation in 1999 \citep{charbonneau.etal1999_DetectionPlanetaryTransits}, the field has blossomed into one of the great scientific frontiers of the 21st century \citep{committeeonexoplanetsciencestrategy.etal2018_ExoplanetScienceStrategy}, with over 6\,000 exoplanet discoveries now having been confirmed,\footnote{\url{https://exoplanetarchive.ipac.caltech.edu}} showing remarkable diversity and allowing us to make significant progress in our understanding of planetary formation and evolution. This diversity has made characterization, particularly of planetary atmospheres, a major field alongside detection \citep{seager.deming2010_ExoplanetAtmospheres, cowan.etal2015_CharacterizingTransitingPlanet}.

Characterizing exoplanet atmospheres can answer important questions about planetary formation and evolution, alongside the potential to uncover the habitability or presence of biosignatures which may even indicate signs of extraterrestrial life \citep{seager.deming2010_ExoplanetAtmospheres}. Traditionally within the field, this characterization of chemical composition and other parameters is performed by iterative physics-based retrievals \citep{barstow.etal2020_ComparisonExoplanetSpectroscopic}. Given the complex nature of the models to be optimized, these methods can be very computationally intensive. With the advent of machine learning in the field, we can achieve significant acceleration of atmospheric characterization \citep{marquez-neila.etal2018_SupervisedMachineLearning, zingales.waldmann2018_ExoGANRetrievingExoplanetary, cobb.etal2019_EnsembleBayesianNeural, fisher.etal2020_InterpretingHighResolutionSpectroscopy, nixon.madhusudhan2020_AssessmentSupervisedMachine, martinez.etal2022_ConvolutionalNeuralNetworks, yip.etal2023_SampleNotSample, lueber.etal2025_NearinstantaneousAtmosphericRetrievals, tahseen.etal2025_ExtremeLearningMachines}.

However, we do not see the proliferation of these techniques across the field; while this is in part simply due to the established nature of existing practices, the issue lies also with two other main shortcomings of machine-learning-based approaches. The first is the availability of data, as machine learning model accuracy is largely dependent on the amount of data used for training \citep{figueroa.etal2012_PredictingSampleSize}, and the largest homogeneously reduced samples of exoplanet transmission spectra currently number only in the tens \citep{tsiaras.etal2018_PopulationStudyGaseousa, edwards.etal2023_ExploringAbilityHST}, far too few to train these tools using real observations. While machine learning tools can be trained using simulated data to predict on real data \citep{lueber.etal2025_NearinstantaneousAtmosphericRetrievals}, the way the training set is generated and sampled shapes what the network learns \citep{fisher.heng2022_HowWeOptimally}, and ensuring that this does not imprint bias onto the trained network is challenging. Determining to what extent bias has been successfully mitigated, and how reasonable and physical predictions are likely to be, is then a high priority. This brings us to the second shortcoming of machine learning techniques: their complex nature renders them `\textit{black-boxes}', whereby obtaining an understandable explanation for their predictions is increasingly challenging \citep{molnarInterpretableMachineLearning2018}. It is this crucial frontier in which the work done with machine learning often falls short not only in this field \citep{marquez-neila.etal2018_SupervisedMachineLearning, 10.1093/rasti/rzag002} but across many others \citep{weber.etal2024_XAITroublea}.

It is important that we are explicit about what constitutes a \textit{black-box} in this context. Popularly, the term is used to refer to machine-learning-based approaches; however, the same \textit{black-box} description applies, if less obviously, to traditional retrievals. A Bayesian retrieval is physics-based in the sense that its forward model, which maps atmospheric parameters to a spectrum, encodes our physical understanding of radiative transfer, chemistry and opacity. The retrieval itself, however, solves the inverse problem: it maps an observed spectrum to a posterior over those parameters. That inverse mapping is not explicitly defined within the Bayesian retrieval codebase. It emerges from the interaction of the forward model with the priors, the likelihood and noise model, and the behaviour of the sampler across a high-dimensional and often degenerate parameter space. Knowing the physics of the forward model does not by itself tell us which wavelength bins are important for a given constraint, or how the posterior would respond if a particular spectral feature\footnote{Here it is necessary to disambiguate between conflicting terminology from machine learning and exoplanetary science. In standard machine learning terminology, an input feature of the model is a single wavelength bin of the spectrum. In exoplanetary science terminology, a spectral feature is a region of the spectrum indicative of a physical effect, as in `between 4 and 5\,\um{} there is a distinct phosphine \textit{spectral feature}', and so spans several adjacent wavelength bins. Where we can, we call input features wavelength bins, and we qualify the word feature wherever the context does not make its meaning clear. A feature named after a molecule, such as the $\mathrm{SO_2}$ feature, is always a spectral feature. For established terminology such as `\textit{feature} importance', which refers to input features, we follow machine learning convention.} were slightly stronger or weaker, or whether degeneracies between parameters are shaping the result. In this sense the input-to-output mapping of a traditional retrieval is also a black-box, and questions of feature importance are as relevant to it as to a machine learning model \citep{welbanks.etal2023_ApplicationBayesianLeaveoneout}. This is also what makes a Bayesian retrieval a useful test bed for an interpretability method: because its forward model is physical, we have independent expectations of which spectral features should matter, against which explanation methods can be compared, while its inverse mapping is opaque enough that those explanations are not trivially known.\Fig{sec:metric}{Section} explores how we can utilize this in more detail.

Given that, logically, more simplistic models can be easier to interpret than complex ones, there is said to be a trade-off between accuracy and interpretability \citep{lundbergUnifiedApproachInterpreting2017a}. A simpler approximation of a complex system is both more intuitive and less representative by definition. Better understanding of how complex black-box models, such as the ones that are adept at solving the atmospheric retrieval problem, reach their predictions can allow us to inspect for biases in both the data and the model, which is especially important given the simulated nature of the data,\footnote{Even when considering Bayesian retrieval, forward models drawn as part of the retrieval constitute simulated data, and therefore any caveats that come with inference on simulated data are equally applicable here as to machine learning models trained using simulated data.} since simulated training data can carry the assumptions of the forward model into the predictions. It is crucial to verify that the model is actually making predictions on the basis of known, physically plausible spectral features, or conversely, in the unlikely case, uncover unexpected relationships which may lead to new domain knowledge. Being able to trace model decisions back to spectral features in the data is key to there being any lasting scientific value behind the method, and can ultimately help improve the usefulness, reliability and ease of adoption of machine learning techniques.

The common machine learning approach to interpretability is to obscure, permute or perturb aspects of the observation in order to determine the relevance of these features to the prediction (known as \textit{feature importance}; \citealt{molnarInterpretableMachineLearning2018}). However, in problems in which the observations themselves are highly correlated, by naively modifying individual wavelength bin values in an observation it is very easy to arrive at a spectrum which is highly non-physical and therefore not within the learned parameter space of the model \citep{maseExplainingBlackBox2020, ng.etal2025_CausalSHAPFeature}. Atmospheric retrieval represents one such example of these correlated problems, as there are very specific ways in which each wavelength bin in an observation is related to the others. At small scales this is due to the effects of spectral line broadening \citep{guest.etal2024_PredictingRotationalDependence}, which remains an ongoing area of development for our understanding of spectral observations \citep{buldyreva.etal2025_SemiclassicalEstimatesPressureinduced}. At less granular scales, this is due to the overall complex structure of molecular transmission as governed by quantum transitions. As a result, out-of-the-box feature importance methods, which typically treat features as independent, can give misleading attributions when inputs are strongly correlated \citep{aas.etal2021_ExplainingIndividualPredictions, maseExplainingBlackBox2020}, and even exact Shapley values are costly to compute and can themselves mislead \citep{huang.marques-silva2023_InadequacyShapleyValues}. Despite this, methods such as SHapley Additive exPlanations (SHAP; \citealt{lundbergUnifiedApproachInterpreting2017a}) and work such as \citet{yipPeekingBlackBox2021a}, which also rely on obfuscation of randomly selected subsets of features, continue to see popularity, in part due to their ease of implementation. In this paper we detail the shortcomings of these types of methods in this application and, in light of this, propose Gaussian perturbation, a new perturbation approach specifically tailored for highly correlated spectral data.

Gaussian perturbation is designed to mitigate the issues with out-of-the-box feature importance methods when used for interpreting machine-learning-based retrievals, handling highly correlated spectral data by generating modified spectra which are instead governed by physically motivated relations. This makes the method much more stable in situations where the response to naive, single-bin perturbation can be noisy and much less smooth. To complement this, the technique does not rely on an accurately constrained posterior, and has a latency which scales with the inference time of the model.

\section{Background}
We assume the reader has familiarity with transit spectroscopy methods; however, for those from a machine learning background, we include a short overview in\Fig{sec:retrieval_background}{Appendix}.

\subsection{Inference from exoplanet observations}
\label{sec:inference_from_exoplanet_obs}

For the purpose of dispelling any preconception (which we acknowledge is disproportionately widespread) that the reader may have regarding the provenance behind machine learning techniques, we first outline the current state of non-machine-learning techniques for atmospheric retrievals. We do this not to present machine learning techniques as a comprehensive replacement for traditional retrievals but to motivate additional diversification of our methodological toolbox, specifically with low-latency strategies suited to the big data era.

There are currently many options available for sampling algorithms, including those that use Markov chain Monte Carlo (MCMC; \citealt{foreman-mackey.etal2013_EmceeMCMCHammer}), those that use nested sampling \citep{speagle2020_DynestyDynamicNested, feroz.etal2009_MultiNestEfficientRobust}, and other more naive sampling methods such as grid sampling. Samplers differ in how reliably they recover the posterior distribution, particularly in the event of strong degeneracies, where there are multimodal likelihood surfaces that random-walk MCMC methods can struggle to explore \citep{feroz.etal2009_MultiNestEfficientRobust, speagle2020_DynestyDynamicNested}. Compounding this, there is an increasing menagerie of separate retrieval implementations \citep{macdonald.batalha2023_CatalogueExoplanetAtmospheric} for the community to choose from, which offer an array of different, sometimes even conflicting, solutions for the same data \citep{barstow.etal2020_ComparisonExoplanetSpectroscopic}.\footnote{See \citet{barstow.etal2016_CONSISTENTRETRIEVALANALYSIS} and \citet{pinhas.etal2019_H2OAbundancesCloud} for two separate interpretations of the same set of hot-Jupiter data from \citet{sing.etal2016_ContinuumClearCloudy}.} This is before we even introduce the effects of different observations or instrumentation \citep{welbanks.etal2023_ApplicationBayesianLeaveoneout}.

This resulting contradiction proliferates the landscape in which authors are compelled to report the results of several separate retrieval methods in order to validate the conclusions of their study \citep[see][as examples of this]{lustig-yaeger.etal2023_JWSTTransmissionSpectrum, dyrekSO2SilicateClouds2024}. The iterative nature of retrieval methods already makes for a very computationally intensive analysis, limiting their use as datasets expand, and the emergence of multi-method validation only compounds this problem. Despite this harsh critique, until very recently these methods were our only tool for handling observational data of this nature, and they remain the most reliable. They therefore remain the default tool, especially in small studies where scaling imposes much less of a constraint.

\subsection{Machine learning retrievals}
As we move into the big data era, the number of studies where the latency of traditional retrievals will become more problematic is only set to rise, and the supplementation of traditional Bayesian retrievals with machine learning in order to minimize the computational cost can help us to prepare for the inevitable challenges this raises, given the rate of increase in data acquisition \citep{martinez.etal2022_ConvolutionalNeuralNetworks, yip.etal2022_ESAArielDataChallenge}.

There are two applications for machine learning in this context. The first is surrogate modelling: replacing the computationally intensive atmospheric models, either in part or in full, with a pre-trained component \citep{tahseen.etal2024_Enhancing3DPlanetary, 10.1093/mnras/staf297}. The second is solving the inverse problem: replacing the Bayesian sampling entirely with a learned representation of the parameter space in order to amortize\footnote{Amortization: \textit{The front-loading of an ongoing operating cost into a high one-time initial cost.}} the problem, offloading the simulation of new spectra into a precomputed training dataset of example spectra. This paper focuses on interpreting the latter of these two approaches, but we note that this methodology can be utilized across a plethora of adjacent regression problems. By using these machine learning approaches to solve the inverse problem directly (inferring chemical abundances directly from reduced spectra) and offsetting the computational expense of inference into a one-time training expense, the trained pipeline can be made comparatively lightweight. Existing work has proven the effectiveness of such techniques in this field, such as \citet{martinez.etal2022_ConvolutionalNeuralNetworks}, \citet{yip.etal2023_SampleNotSample} and \citet{lueber.etal2025_NearinstantaneousAtmosphericRetrievals} \citep[for a full list see][]{macdonald.batalha2023_CatalogueExoplanetAtmospheric}.

\subsection{Interpretability}
\label{sec:interpretability}
By introducing any black-box component into a retrieval pipeline, it becomes challenging to trace the predicted abundances back to spectral features in the observed data. The most important aspect that we want to untangle is the feature importance, as introduced in\Fig{sec:intro}{Section}, which is the relative impact of each input wavelength bin in our observed data on the results of our retrieval.

\subsubsection{Global vs local importance}
\label{sec:global_local}
There are two types of feature importance which are relevant for this work, global feature importance and local feature importance. Global feature importance is independent of any observation. Global feature importance allows for statements such as `across the atmospheres this retrieval can return, its estimate of the $\mathrm{CO_2}$ abundance depends most on the wavelength bins around 4.3\,\um{}', which tells us whether the retrieval relies on physically plausible regions of the spectrum at all, independent of any one observation. Local feature importance is relative to a prediction on one specific observation. Local feature importance allows for statements to be made such as `the most important wavelength bin in \textit{this} retrieval for the abundance of $\mathrm{CO_2}$ is 4.3\,\um{}; we set a lower bound of $x$ for $\mathrm{CO_2}$ abundance \textit{given that} the lower bound of the uncertainty on the 4.3\,\um{} wavelength bin is within $y$', which is very important for realistic classification of low-confidence detections of molecules with very narrow wavelength signatures, where the whole retrieval may hinge on very few input features.

\subsubsection{Data-driven vs model-driven importance}
\label{sec:data_model_driven}

A second distinction concerns what the importance describes. Data-driven importance describes the information that the data carry about each parameter given the forward model, i.e. which wavelengths a retrieval faithful to the forward model ought to use. It can be computed from forward models alone, without reference to the retrieval. Model-driven importance describes the retrieval itself, i.e. which wavelengths it actually uses, and can only be obtained by probing the retrieval, for example by perturbing its input. For an accurate retrieval the two agree, and where they differ it is the model-driven importance that shows what the retrieval is doing.

Combining the two distinctions gives four kinds of importance. The contribution functions of a forward model, and the global metric of\Fig{sec:global_metric}{Section}, are global and data-driven, while the importance of a retrieval based on the internal working of the retrieval code, such as its weights in the case of a machine learning model, is global and model-driven. The local metric of\Fig{sec:local_metric}{Section}, computed from the posterior for one observation, is local and data-driven, and the perturbation importance of one observation, the subject of this work, is local and model-driven. For a Bayesian retrieval the local data-driven importance can be computed, which gives a reference against which to test the perturbation importance\Figp{sec:metric_comparison}{Section}. For a machine learning retrieval without a calibrated posterior it cannot, and only model-driven importance is available.

\subsubsection{Existing methods}

Local feature importance can be obtained for traditional retrievals \citep{welbanks.etal2023_ApplicationBayesianLeaveoneout} by means of leave-one-out Bayesian methods. As an aside, the existence of such tools for Bayesian retrievals motivates this work twofold, first by underlining that the question of `which wavelengths drive a given constraint' within a retrieval is not answered by the physics of the forward model alone, and secondly by showing that this information is relevant and sought after. Leave-one-out cannot be explicitly applied to most machine learning networks as they are trained to accept a fixed number of input bins, so instead the target bin is modified or \textit{`perturbed'} so as to mask the information that it would otherwise have provided to the retrieval. Perturbation feature importance has been explored for machine learning retrievals in this field, for example in \citet{yipPeekingBlackBox2021a}. This work perturbs random groups of wavelength bins from their uncertainties, and then uses the respective change in the prediction over many repetitions and test spectra as an unsigned global sensitivity map. This `perturbation' methodology therefore presents us with a logical starting point for our exploration into improved interpretability; first we should understand the way such techniques work, and their limitations.

\subsection{Retrievals as black-box functions}

Any retrieval can be described as a function $f$ such that the predicted parameters $\hat{\vect{y}}_\star=f(\vect{x}_{\star})$, the mean of the retrieved posterior\Figp{sec:formalising}{Section}, where $\vect{x}_{\star}$ is a given observation. This is true for either machine learning or Bayesian retrievals. Perturbation-style interpretability is simply an exercise in mapping a change $\delta\vect{x}_{\star}$ on to the respective change $\delta\hat{\vect{y}}_\star$ in a human-readable manner, and therefore agnostic of the mechanisms of the function $f$.

For a Bayesian retrieval, the physics is contained in the forward model; however, $f$ is the whole inference procedure: the forward model together with the priors, the likelihood and the sampler. As discussed in\Fig{sec:intro}{Section}, this makes it a black-box, but one with independent physical expectations against which an interpretability method can be tested. Since the focus of this paper is interpretability, and not machine learning models, in an attempt to convince the reader of the authenticity of the interpretability methods that we present, we first interpret retrievals performed by the Bayesian retrieval framework \textsc{TauREx-III} \citep{al-refaie.etal2021_TauRExIIIFast} in lieu of a machine learning model. This resolves any concern that an unexpected explanation reflects a flaw in a poorly trained network, and allows us to focus on the effects of the interpretability frameworks with the confidence that our retrieval mechanism is both well-established and physically motivated. We then apply the method to a machine learning retrieval in\Fig{sec:ml}{Section}.

\subsection{Formalizing the retrieval problem}
\label{sec:formalising}

We denote the black-box retrieval as a function $f$, which relies on a set $\mat{X}$ of $N$ data points (observations) which are pairs of the form ($\vect{x}_n$, $\vect{y}_n$), $\forall n \in [1,\dots,N]$, where $\vect{x}_n =[x_{n,1},x_{n,2},\dots,x_{n,M}] \in \mathbb{R}^{1 \times M}$ corresponds to a spectrum consisting of $M$ wavelength bins $x_{n,m}\in \mathbb{R}$, $\forall m\in[1,\dots,M]$ and $\vect{y}_n = [\log(\mathrm{H_2O}), \log(\mathrm{CO_2}), \log(\mathrm{SO_2}), \dots] \in \mathbb{R}^{1 \times P}$ is a vector representing the $P$ atmospheric parameters (here molecular abundance of each atmospheric gas) to be predicted. For a machine learning retrieval this set $\mat{X}$ is its training data, and for a Bayesian retrieval it is the set of forward models evaluated during sampling. We denote the \textit{unknown} distribution from which $\vect{x}_n$ is sampled as $\mathcal{D}$, the distribution of spectra generated by the assumed forward-model parameter distribution. Given an observed spectrum $\vect{x}_\star$ as input, the model can output the predicted atmospheric parameters as $\hat{\vect{y}}_\star=f(\vect{x}_{\star})$. Conversely, a forward model can be viewed as a function $g$ that generates a spectrum using as inputs the atmospheric abundances, $\vect{x}_\star=g(\vect{y}_\star)$. In other words, the retrieval function $f$ serves as an approximation of the inverse of the forward model $g$, i.e. $f \approx g^{-1}$. A retrieval returns a posterior distribution over the atmospheric parameters rather than a point estimate. Throughout this paper the output of $f$ is the mean of that posterior, so that $f$ maps each observed spectrum to a single vector of parameters, $\hat{\vect{y}}_\star = f(\vect{x}_\star)$, and not to the posterior itself.

For \textsc{TauREx-III} this is the mean over the nested-sampling samples, each weighted by its posterior weight, and for the convolutional neural network (CNN) retrieval of\Fig{sec:ml}{Section} it is the unweighted mean of the network's predictions over a fixed set of 1\,000 Monte Carlo dropout masks. Where we quote another summary of a posterior, such as its median, we say so. For readability, we drop the index $n$ for the rest of the paper when it is clear that we refer to a specific sample.

\section{Ceteris paribus profiles}
\label{sec:cetpar}

\begin{figure}
    \centering
    \includegraphics[width=0.45\textwidth]{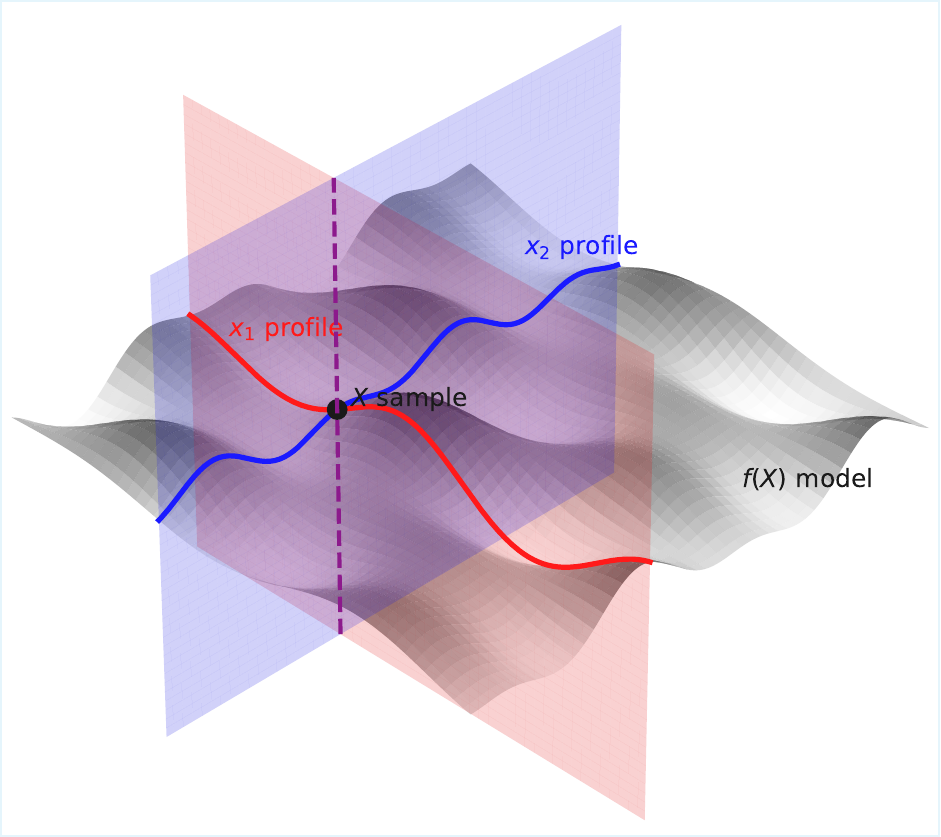}
    \caption{Ceteris paribus profile visualization for the simplest two-dimensional case. For a model $\hat{y}=f(\vect{x})$ with input features (or wavelengths) $x_1$ and $x_2$, which predicts the abundance of one molecular species $y$, the ceteris paribus profiles for both input features of a specific given observation $\vect{x}_n$ are visualized.}
    \label{fig:cetpar_intro}
\end{figure}

\textit{Ceteris paribus}\footnote{Latin: `other things (being) equal'.} profiles are one of the simplest forms of explanation for black-box functions. They are used to characterize how the model responds to variations in input values. The premise of ceteris paribus represents the simplest experiment one can perform to characterize a system: vary one input feature while keeping all others constant, and measure the resulting change in the output (in our case, the output is the molecular abundance of a single species). Despite this naive approach, ceteris paribus profiles allow for a visualization of the complex curvature of the model function's surface, projected into the dimension representing a single input feature, with all other input features fixed at a set value \citep{burzykowski_10CeterisparibusProfiles}. This can provide a very powerful description of the local parameter space around the fixed features. For a one-dimensional problem, this can be a complete global picture of the system. For a two-dimensional system as in\fig{fig:cetpar_intro}, this is a good approximation of the immediate parameter space surrounding the observation. As dimensionality increases, however, the parameter space becomes orders of magnitude more complex, so that at high dimensions, this method is substantially less informative.

\subsection{Mathematical framework of naive perturbation}

Formally, for a given observation $\vect{x}_\star$, we calculate the response of the model to a change in the value $x_{\star,m}$ of wavelength bin $m$ by retrieving abundances over a set of modified observations. This analysis is only meaningful given that we limit the change in the value $x_{\star,m}$ to the range within which it would reasonably be expected to lie, so let us set minimum and maximum values as $\alpha_{\max} \times$ observational error $\sigma_m$, where $\alpha_{\max}$ is the perturbation magnitude scaling parameter. In the \mbox{HAT-P-26}~b experiments we use $\alpha_{\max} = 3$, and for HD~209458~b $\alpha_{\max} = 2$\Figp{sec:maths}{Section}.

\begin{equation}
   z_m^{\min} = x_{\star,m} - \alpha_{\max}\,\sigma_m, \quad z_m^{\max} = x_{\star,m} + \alpha_{\max}\,\sigma_m.
\end{equation}
 Using this context, a linear grid of $K$ values between these, $\vect{z}_m \in \mathbb{R}^{1\times K}$, can be constructed:
\begin{equation}
    z_{m,k}=z_m^{\min}+\frac{(k-1)(z_m^{\max}-z_m^{\min})}{K-1},\quad\forall k \in \left[1,2,3,\dots,K\right].
\end{equation}
As such we can define, where the operator `$:=$' denotes value assignment, our set of modified observations $\mat{X}^\prime_{m} \in \mathbb{R}^{K\times M}$ where the $m^{\textrm{th}}$ wavelength bin is replaced with each of the intensity values from $\vect{z}_m$ in turn,
\begin{equation}
    \vect{x}^\prime_{m,k} = \vect{x}_\star|_{x_{\star,m}:=z_{m,k}}, \forall z_{m,k} \in \vect{z}_m,
\end{equation}
and \textit{all other values being equal}, i.e. left the same as in the initial observation $\vect{x}_\star$. This set of modified observations is then specific to the observation $\vect{x}_\star$. By subsequently retrieving abundances for all modified observations we can discretely characterize the modified retrieved abundances $\hat{\mat{Y}}^\prime_{m}$ \citep{molnarInterpretableMachineLearning2018} as
\begin{equation}
    \hat{\mat{Y}}^\prime_{m} = f(\mat{X}^\prime_{m}).
\end{equation}
The response profile $\vect{r}_{m,p} \in \mathbb{R}^{K\times 1}$ of molecule $p$ to wavelength bin $m$ is the change in its retrieved abundance over the grid $\vect{z}_m$; for $p = \mathrm{SO_2}$,
\begin{equation}
    \vect{r}_{m,\mathrm{SO_2}} = \hat{\vect{y}}^{\prime}_{m,\mathrm{SO_2}} - \vect{1}_K\,\hat{y}_{\star,\mathrm{SO_2}}.
\end{equation}
We refer to perturbing a single wavelength bin in this way as naive perturbation.

\subsection{Limitations for correlated data}
The main advantages of ceteris paribus are that it is typically fast, model-agnostic, and can work well if feature contributions are independent. The main drawback of ceteris paribus profiles is that in the case of non-independent input feature interactions, some, if not most, of the set of modified observations generated by the method can be non-physical. These out-of-distribution samples prompt unstable model response, which is not useful to learn.

To revive our earlier example\Figp{sec:global_local}{Section}, even if the upper bound of the observational uncertainty on the $4.3\,\um$ wavelength bin is very high, a value this high can only arise in very $\mathrm{CO_2}$-dominated atmospheres, in which case the $4.2\,\um$ and $4.4\,\um$ wavelength bins would also be high. However, since we are exploring the upper limit on the $4.3\,\um$ wavelength bin \textit{only}, the $4.2\,\um$ and $4.4\,\um$ wavelength bin values will not be correspondingly high, and so the modified observations in $\mat{X}^\prime_{m}$ with high $z_{m,k}$ will be non-physical. In other words,
\begin{equation}
    \vect{x}^\prime_{m,k}\nsim\mathcal{D} \quad \text{for large } z_{m,k} - x_{\star,m}.
\end{equation}
This means that the resulting $\vect{r}_{m,p}$ describes how the retrieval responds to spectra outside the range of inputs it is designed for, which need not be how it responds to physical changes in the spectrum.

In data regimes where the points are highly correlated, as in spectral retrieval, ceteris paribus-style methods suffer heavily from the generation of out-of-distribution samples. Sampling-based attribution methods such as SHAP partially mitigate this by drawing replacement values from a background dataset, but mixing wavelength bins from different spectra still rarely yields a coherent spectrum \citep{aas.etal2021_ExplainingIndividualPredictions, maseExplainingBlackBox2020}. Work has been carried out into how to mitigate this \citep{maseExplainingBlackBox2020} and avoid generating out-of-distribution samples in a problem-agnostic way. Likewise, the methodology \cite{yipPeekingBlackBox2021a} partially mitigates this by perturbing sets of wavelength bins together, but these randomly drawn sets will rarely align with correlation structure. Neither of these approaches are optimal given that the specific nature of correlation within spectral samples is very well understood; in this field we can likely do better by applying existing domain-specific knowledge rather than inferring it.

\section{Data}
Data selection for this study was constrained by a strict set of requirements. Firstly, we wanted to look only at papers which provided a published \textsc{TauREx-III} retrieval with parameters and open-access data, so that we could replicate the retrieval with minimal justification required. Secondly, due to the use of Bayesian methods in place of machine learning networks, we needed a retrieval with a low computational cost in order to compute our interpretability metrics iteratively. This means a relatively simple atmosphere without exotic chemistry and with few free parameters. While \textit{Hubble Space Telescope} (\textit{HST}) data are much simpler in this regard, we recognize the importance of demonstrating our methodology with more modern \textit{James Webb Space Telescope} (\textit{JWST}) data as well. As such we consider both \textit{HST} and \textit{JWST} data. A strong transit signature favours planets in the hot Jupiter to warm Neptune range; covering both ends of this range is a useful, though not essential, diversification.

We also needed a strong detection of a molecule which has distinct regions where it is degenerate with other parameters, and regions where it is not, such that we have a clear and obvious benchmark against which to verify that the interpretability framework has identified and characterized regions of importance correctly. We define these as strongly retrieved molecules which have narrow transmission features within the wavelength range of the observation. The molecules involved also need publicly available linelists. Throughout this work we use linelists provided by ExoMol \citep{polyansky.etal2018_ExoMolMolecularLine, azzam.etal2016_ExoMolMolecularLine,bowesman.etal2024_ExoMolLineLists, chubb.etal2021_ExoMolOPDatabaseCrosse, coles.etal2019_ExoMolMolecularLine, mckemmish.etal2019_ExoMolMolecularLine, sousa-silva.etal2015_ExoMolLineLists, underwood.etal2016_ExoMolMolecularLine, yurchenko.etal2022_ExoMolLineLists,yurchenko.etal2024_ExoMolLineLists, yurchenko.etal2026_ExoMolLineLists}.

\subsection{HD~209458~b}
The first dataset used in this paper is the \textit{HST} Wide Field Camera 3 (WFC3) spectrum of HD~209458~b from \cite{tsiaras.etal2018_PopulationStudyGaseousa}, which is used as the benchmark for \textsc{TauREx-III} in \cite{al-refaie.etal2021_TauRExIIIFast}. This observation presents a detection of water, with no strong contribution from any other molecule, and is very lightweight, allowing for an average retrieval time of 36 minutes. We replicate the exact transmission retrieval performed on this data in Section 8.2 of the \cite{al-refaie.etal2021_TauRExIIIFast} benchmark paper, the parameters and priors of which can be found in the appendix in\Fig{tab:hd209_params}{Table}
 and\Fig{tab:hd209_priors}{Table}. Posteriors from the retrieval can be found in the appendix in\Fig{tab:HD209_posteriors}{Table}.

\subsection{\mbox{HAT-P-26}~b}
The second dataset used in this paper is the \textit{JWST} Near Infrared Spectrograph (NIRSpec) G395H spectrum of \mbox{HAT-P-26}~b from \cite{gressier.etal2025_JWSTTSTDREAMSSulfur}. This observation was chosen as it features prominent detections of $\mathrm{CO_2}$ and $\mathrm{SO_2}$, which in the G395H region both exhibit narrow spectral features, as well as some broader features which we can use to explore the resolving power of these interpretability techniques. The paper presents several reduction/retrieval combinations of the observation data; the one which we replicate is the \textsc{transitspectroscopy} \citep{espinoza2022_TransitSpectroscopy} reduction at a resolving power of $R\sim100$, with the \textsc{TauREx-III} retrieval. The Bayesian retrieval of this data is more computationally intensive than that of HD~209458~b, with an average retrieval time of around 6.5 hours, and so we use it primarily for machine learning retrievals and other low-latency statistical exploration, alongside selective Bayesian retrievals to verify claims explored using the HD~209458~b data in higher resolution. Parameters and priors of this retrieval can be found in the appendix in\Fig{tab:hatp26_params}{Table}
 and\Fig{tab:hatp26_priors}{Table}. Posteriors from the retrieval can be found in the appendix in\Fig{tab:HATP_posteriors}{Table}.

\section{Gaussian perturbation}
Ceteris paribus profiling is a very effective tool for data we can be certain are uncorrelated. Exoplanet spectra are not uncorrelated, but the correlation between wavelength bins within them is very well characterized. This means that we could, in principle, interpret some less correlated representation of the spectra using a ceteris paribus technique.

While this is true in isolation, it is important to remember that the whole focus of using these interpretability techniques is increased transparency. There is little value in understanding the model interpretation of some complex reduction of the observed data where there is no clear relation back to the observation. Therefore, a balance needs to be struck where modified observations are more likely to remain consistent with the distribution $\mathcal{D}$, without being defined by an overly complex or arbitrary function.

As a result of this, rather than remove the correlation from the spectra, it is more logical to impose a set of rules governing the selection of wavelengths which are to be modified such that correlations are maintained. The first-order driver of correlation between any two data points in an atmospheric spectrum is wavelength proximity. Correlation is not purely local: a molecule absorbs in several bands, so bins in different bands of the same gas, such as the $\mathrm{SO_2}$ features at 3.65 and $4.04\,\um$, can also be strongly correlated. Such correlations are specific to particular molecules and wavelength regions, whereas adjacent data points are, on average, more correlated than non-adjacent ones, as\fig{fig:correlation} shows for every pair of wavelength bins. This is the result of broad-band absorption by the more abundant molecules. Looking towards more and more exotic atmospheres containing molecules with very narrow spectral signatures, this correlation will decrease, but ultimately still dominates at low resolution. Perturbing adjacent wavelengths together should therefore produce samples that better preserve local spectral structure and are more likely to remain physically plausible.

\subsection{Choice of kernel width}

\begin{figure}
    \centering
    \includegraphics[width=0.48\textwidth]{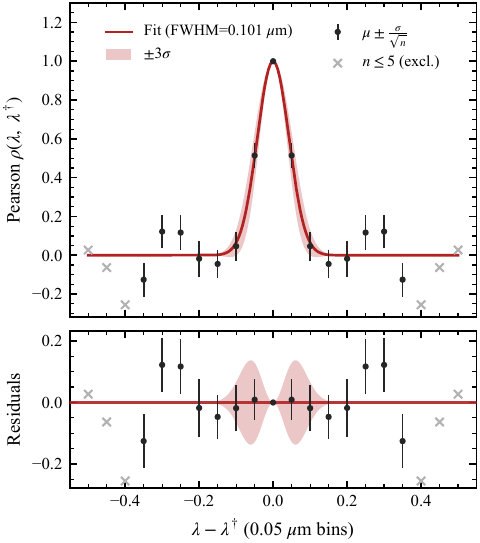}
    \caption{Pearson correlation between all possible pairs of wavelengths $\lambda$ and $\lambda^\dagger$ across the grid of forward models of\fig{fig:correlation}, visualized by the $\lambda - \lambda^\dagger$ distance. The data are binned into $0.05\,\um$ bins. Because of the finite width of the observation there are fewer pairs at large wavelength separations; the number of samples per bin is much higher towards the centre of the plot. To mitigate any effects of this we present error bars here as standard error rather than standard deviation, and bins with five or fewer data points are excluded. The Gaussian fit yields an optimal kernel FWHM of $0.101\,\um$.}
    \label{fig:gaussian_fit}
\end{figure}

The Gaussian technique approaches the non-physical sample generation problem by probing the response of a cluster of adjacent bins defined by a Gaussian mask. This Gaussian assumption holds as an approximation to first-order correlation. It therefore captures only the local part of the correlation, and does not perturb the other bands of a molecule together, since the aim here is to attribute a retrieved parameter to regions of the spectrum rather than to a molecule as a whole. A kernel built from the full correlation matrix of the forward models could include the correlation between bands, but this is out of scope for this work. The optimal width of the Gaussian (defined as constant in the wavelength scale, not the bin scale) is to a certain extent dependent on the target molecule and the binning grid. The use of a wider window will reduce feature importance resolution, whereas a narrower window will generate more non-physical samples and reduce the power of the technique. Therefore, the optimal Gaussian is one that matches the scale of the spectral features of interest in the observation. A single molecule, however, can have spectral features of many widths, from narrow bands to broad absorption, and no one width suits all of them. In that case the width can be chosen for the specific molecule and features of interest within the observation wavelength range. The analysis could be repeated over several widths; however, this multiplies the number of retrievals and increases the computational cost significantly, so it is not recommended as common practice outside exceptional situations where a specific degeneracy or instrumental quirk is known beforehand to raise issues. Preliminary testing suggested that the technique performs best with a Gaussian of full width at half maximum (FWHM) $0.1\,\um$. The selected kernel width is appropriate for the spectral configurations considered here and should not be assumed optimal for other instruments, resolutions or atmospheric regimes. We assessed the Pearson correlation between all wavelengths across the grid of forward models of\fig{fig:correlation} and applied a fit to the resulting data points as in\fig{fig:gaussian_fit}, to find an optimal Gaussian FWHM of $0.101\,\um$, consistent with our preliminary findings. Best practice with the technique would involve validating the assumption using this type of analysis on some sampling of the parameter space for each set of priors or binning grid used. It is important to note that this assumption holds better at some wavelengths than at others, so we have to be wary of specific wavelengths which are receiving poor representation under this static Gaussian assumption.

To limit deviations from the observed spectrum, the maximum perturbation magnitude is set using the instrument noise profile for the target observation. This constrains the modified observations to remain close to the observed data, although it does not guarantee their physical consistency.

\subsection{Mathematical framework of Gaussian perturbation}
\label{sec:maths}

We can formalize this methodology: first we lay out the known nature of the relationships in our data. As is common in physics notation, we can define a neighbourhood $\mathcal{N}(d,m)$ of wavelength bin $m$: the set of other bins whose wavelengths lie within some small distance $d$ of $\lambda_m$,
\begin{equation}
    \mathcal{N}(d,m) = \{\, i \neq m : |\lambda_i - \lambda_m| < d \,\}, \quad \ell_m = |\mathcal{N}(d,m)|,
\end{equation}
where $\ell_m$, the number of bins in the neighbourhood, varies with $m$ on a non-uniform wavelength grid. Writing $\rho_{mi} = \rho(\mat{X}_{:,m}, \mat{X}_{:,i})$ for the Pearson correlation between bins $m$ and $i$ across the $N$ spectra, this notation allows us to define the relationship where neighbouring points are more positively correlated than non-neighbouring points,
\begin{equation}
\begin{aligned}
0 \lessapprox{}& \frac{1}{M}\sum_{m=1}^{M}\frac{1}{M-1-\ell_m}\sum_{i \notin \mathcal{N}(d,m)\cup\{m\}} \rho_{mi}\\
<{}& \frac{1}{M}\sum_{m=1}^{M}\frac{1}{\ell_m}\sum_{i \in \mathcal{N}(d,m)} \rho_{mi}.
\end{aligned}
\end{equation} For example, in\fig{fig:correlation}, we see that there are few cases of strong negative correlation in the data.

Our experiments\figp{fig:gaussian_fit} can go further to show that the magnitude of correlation between wavelength bins is inversely related to the distance between the bins. In practice, we can define this as a windowing function $\vect{w}_{\text{gauss.}}(\varepsilon,m) \in \mathbb{R}^{1\times M}$ with a Gaussian weighting of FWHM $\varepsilon$ in wavelength,
\begin{equation}
    w_{\text{gauss.}}(\varepsilon,m)_j=\exp\left(-\frac{4 \ln 2 \,(\lambda_j - \lambda_m)^2}{\varepsilon^2}\right), \quad j = 1,\dots,M.
\end{equation}
With the data correlation constrained in this way, it becomes far easier to generate samples which are consistent with the prior distribution. By applying a windowing function that approximates local spectral correlations, we can `diffuse' the perturbation across neighbouring wavelength bins, generating samples that better preserve the expected spectral structure and are therefore less likely to be out-of-distribution.

Given a known instrument noise $\boldsymbol{\sigma}$, the perturbation around wavelength bin $m$ is set by a preset grid of perturbation magnitudes $\boldsymbol{\xi} \in \mathbb{R}^{K\times 1}$, in units of the noise $\sigma_m$ of that bin (which we quote e.g. $\xi = 3\sigma$),
\begin{equation}
   \boldsymbol{\xi} = \begin{bmatrix} 3 \\ 0\\ -3 \end{bmatrix},
\end{equation}
shown here for $\boldsymbol{\xi}$ linearly spaced on $[-\alpha_{\max}, \alpha_{\max}]$ with $\alpha_{\max} = 3$ and $K = 3$. The number of \textsc{TauREx-III} retrievals is limited by computational cost, so this three-point grid is the one used for\fig{fig:observation},\fig{fig:marginal_posteriors} and the full posteriors of\Fig{sec:full_posteriors}{Appendix}, while the HD~209458~b results in\fig{fig:perturb_showcase_figure},\fig{fig:gradient_local_metric},\fig{fig:final_scatterplot} and\fig{fig:final_scatterplot_all} use the non-linear grid $[0, \pm(0.5, 1, 2)]\sigma$. For the machine learning retrieval in\fig{fig:hatp_corner_co2_so2}, which is far cheaper to evaluate, we use a linear grid with $K = 101$ and $\alpha_{\max} = 3$, a step of $0.06\sigma$.

Combining these factors the perturbation response is then calculated. For a given observation $\vect{x}_\star$, the response $\mat{R}_{m} \in \mathbb{R}^{K\times P}$ of all $P$ molecules to the region centred on wavelength $m$ can be calculated by retrieving the set of abundances as $\hat{\mat{Y}}^\prime_{m} = f(\mat{X}^\prime_{m})$ over the subset of modified observations
\begin{equation}
    \mat{X}^\prime_{m} = \vect{1}_K\vect{x}_\star + \boldsymbol{\xi}\sigma_m\: \vect{w}_{\text{gauss.}}(\varepsilon,m),
\end{equation}
and comparing this against the retrieved abundances $\hat{\vect{y}}_\star$ for the zero-perturbation observation where $\hat{\vect{y}}_\star = f(\vect{x}_{\star})$, to get the perturbation response
\begin{equation}
    \mat{R}_{m} = \hat{\mat{Y}}^\prime_{m} - \vect{1}_{K}\hat{\vect{y}}_\star.
\end{equation}
Its column for molecule $p$, $\vect{r}_{m,p}$, is the response profile defined above for ceteris paribus profiles. Note that for $\xi_k = 0$, $\mat{X}^\prime_m$ is invariant under $m$, so computational shortcuts can be taken here.

The whole perturbation is scaled by the noise $\sigma_m$ of the centre bin, so that a neighbouring bin $j$ is perturbed by $\xi_k\,w_j\,\sigma_m/\sigma_j$ in units of its own noise. An alternative is to scale each bin by its own noise, $\xi_k\,\sigma_j\,w_j$, which keeps every bin within the same number of its own standard deviations but makes the perturbation follow the bin-to-bin variation of the noise rather than the smooth shape of a change in the spectrum. The two agree only where neighbouring bins have similar uncertainties. Within the half maximum of the kernel, the noise of a neighbouring bin differs from that of the centre bin by 9.0 per cent on average for HD~209458~b and 10.0 per cent for \mbox{HAT-P-26}~b, and by less than 5 per cent around the $\mathrm{SO_2}$ and $\mathrm{CO_2}$ bands targeted in\Fig{sec:hatp}{Section}. The largest difference is at the blue edge of the G395H spectrum, where the first bin, at $2.770\,\um$, is 54.5 per cent noisier than its neighbour at $2.799\,\um$: a kernel centred on the first bin with $\xi = 3\sigma$ moves the neighbour by $3.7\sigma_j$ rather than $2.4\sigma_j$, beyond the $\pm3\sigma$ range of the grid. Where the noise changes sharply between neighbouring bins, a bin with much lower noise than the centre bin is perturbed by correspondingly more of its own standard deviations, and the perturbed spectra can leave the distribution the method is designed to stay within. For such data consider instead if each bin should be scaled by its own noise. For this work, we present the simplest version of the technique, and retain the scale of the centre bin.

\section{Dimension reduction of naive and Gaussian samples}
\label{sec:dim_reduction}

\begin{figure}
    \centering
    \includegraphics[width=0.48\textwidth]{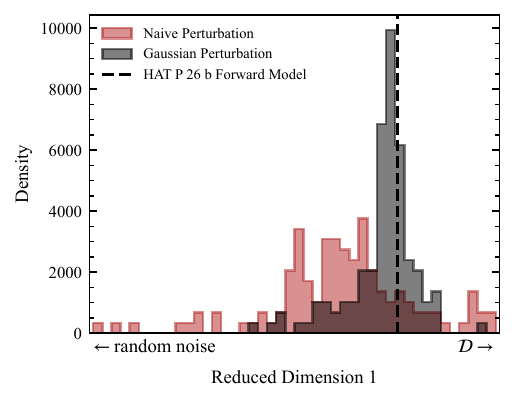}
    \caption{Each spectrum on the \mbox{HAT-P-26}~b wavelength grid is mapped to a single score by a small convolutional classifier, trained to separate forward models drawn from the \mbox{HAT-P-26}~b priors (score 1) from randomly sampled noise (score 0;\Fig{sec:dim_reduction}{Section}). The set of forward models drawn from \mbox{HAT-P-26}~b priors\Figp{tab:hatp26_priors}{Table} occupy the right-hand side of the parameter space, whereas `observations' comprised of randomly sampled noise occupy the left-hand side. A forward model generated from the posterior means retrieved for the \mbox{HAT-P-26}~b observation is shown by the dashed line. The set of modified spectra produced by Gaussian perturbation of the \mbox{HAT-P-26}~b forward model is shown as a histogram in black, and the set of modified spectra produced by naive perturbation is shown in red. `Gaussian' refers to the shape of the perturbation kernel in wavelength, not to the distribution of the modified spectra, each of which is a fixed $\pm1\sigma$ perturbation centred on one bin, so the histograms need not be Gaussian. Gaussian perturbation produces modified samples that cluster more closely around the target observation than those of naive perturbation.}
    \label{fig:dim_reduced_population}
\end{figure}

Logically it follows that the Gaussian method of data perturbation produces fewer spurious samples than a naive implementation, but it can also be visualized in a lower-dimensional representation for clarity. For this demonstration we use the \mbox{HAT-P-26}~b data. By compressing the 60 wavelength bins in the observation into a single dimension using a convolutional reduction\footnote{While we recognize the limitations and potential issues associated with dimensionality reduction techniques, they are employed here solely as a means of providing additional visual clarity. We chose to use the convolutional reduction of the data despite the fact that its representation can be influenced by parameter choices and data density, as it is not our intention to provide any quantitative evidence from the embedding itself.} of a set of forward models drawn from \mbox{HAT-P-26}~b priors, we can compare a representation of the spectra on a scale of those which are `in-distribution' and `out-of-distribution'\figp{fig:dim_reduced_population}. The convolutional reduction describes an approximation to the non-linear transformation that maps the observation on to a single-variable scale with random noise at $0$ and in-distribution samples from the prior at $1$.\footnote{The classifier has three convolutional layers ($32$, $32$ and $64$ filters, kernel width of 3 bins, max pooling after the
second), global average pooling and two dense layers ($32$ units with dropout $0.2$, then $1$ unit). Each wavelength bin
is scaled to $[0, 1]$ by its minimum and maximum across the forward models. It is trained with a binary cross-entropy
loss for $8$ epochs (Adam, learning rate $10^{-3}$, batch size $64$) on $80$ per cent of the set, and classifies all of the
remaining $20$ per cent correctly.} We can expect the retrieval performance for spectra that are closer to the `in-distribution' end of the scale to be stable. However, for spectra that appear closer to the `out-of-distribution' random noise side, the performance of the retrieval is completely uncharacterized, as these represent spectra which are either (a) formed by combinations of abundances that are outside the prior constraints, or (b) not physically plausible spectra. Either way, we cannot expect to gain any meaningful information about the retrieval response from these.

We cannot use the raw \mbox{HAT-P-26}~b observation spectrum in this reduction space, as this will contain subtle instrumental and reduction artefacts which are not present in the forward model grid used to construct it. Therefore, we construct a forward model at the posterior means retrieved for the raw \mbox{HAT-P-26}~b observation spectrum and use this forward model in place of the actual observation data here. This does not affect the comparison, which concerns where the perturbed spectra lie relative to the spectrum they are generated from: Gaussian and naive perturbation start from the same forward model, and therefore any artefacts in a real spectrum would be common to both. We perturb it by $\xi = \pm1\sigma$, centring the kernel on each of the 60 bins in turn, to give $120$ Gaussian and
$120$ naive modified spectra. When the kernel is centred on or near the first or last bin, it is truncated by the edge of the spectrum, so only the bins on one side are perturbed, while the centre bin is still perturbed by the full $\xi\sigma_m$.

The simulated \mbox{HAT-P-26}~b observation appears closer to the `in-distribution' end of the reduction scale, as it is in accordance with the retrieval priors and the physical restrictions, and therefore internally consistent with what we would expect from a spectral observation. As the naively perturbed spectra are generated by perturbing individual wavelength bins not in accordance with any physical model, they are not necessarily required to be internally consistent in line with the set of correlations that we would expect from a physical spectral observation. It is very likely that some naively perturbed spectra exhibit spectral features which are non-physical, which are represented here as samples that are towards the random noise end of the reduction scale.

Conversely, Gaussian perturbation imposes a smooth, locally correlated structure on the modified spectra. As illustrated in\fig{fig:dim_reduced_population}, these samples lie closer to the reference forward-model distribution in the reduced representation, suggesting improved coverage of the model response near the target observation.

\section{Application to \texorpdfstring{\textsc{TauREx-III}}{TauREx-III}}
\label{sec:hd209_methodology}
We apply Gaussian perturbation to the \textsc{TauREx-III} retrieval of HD~209458~b. A kernel of FWHM $\varepsilon = 0.1\,\um$ is centred on each of the 18 wavelength bins of the WFC3 spectrum in turn, at each of the six non-zero magnitudes of the grid $\boldsymbol{\xi} = [0, \pm(0.5, 1, 2)]\sigma$, and each perturbed spectrum is retrieved with the same priors and settings as the unperturbed one\Figp{tab:hd209_priors}{Table}, with 108 perturbed retrievals and the unperturbed one shared by every bin. From each retrieval we take the posterior mean of the log abundance of every gas, and the response $\vect{r}_{m,p}$ of gas $p$ to bin $m$ is its shift from the unperturbed posterior mean at each magnitude. For comparison, we also perturb each wavelength bin individually by the same magnitudes, i.e. naive perturbation\Figp{sec:cetpar}{Section}, in a further 108 retrievals that share the same unperturbed one.

\subsection{Visualization}
\label{sec:visualisation}
\begin{figure}
    \centering
    \includegraphics[width=0.48\textwidth]{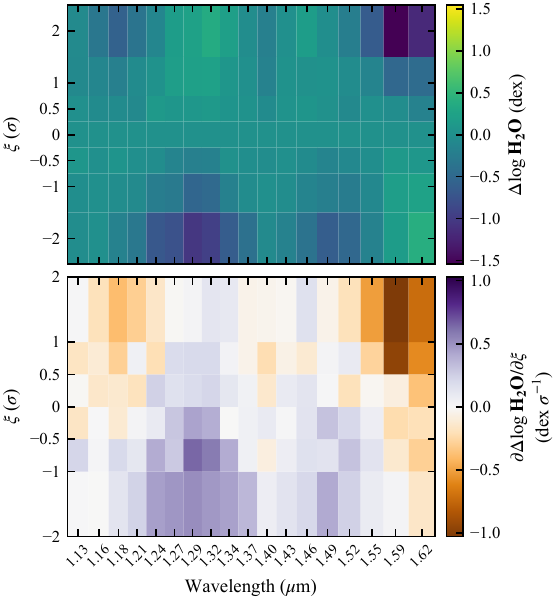}
    \caption{The upper panel shows the shift (in dex) of the posterior mean of $\log\mathrm{H_2O}$ in response to the applied perturbation kernel centred on each wavelength, for each perturbation magnitude $\xi$. The sign of $\xi$ gives the direction of the perturbation, positive raising the transit depth under the kernel and negative lowering it, and its absolute value gives the size, in units of $\sigma_m$. The response need not be symmetric in $\xi$\Figp{sec:visualisation}{Section}. The lower panel shows the derivative of this, the rate of change of the retrieved abundance of $\mathrm{H_2O}$ with perturbation magnitude, between neighbouring magnitudes; its mean down each column is the importance of that wavelength.}
    \label{fig:perturb_showcase_figure}

\end{figure}
Given the non-linearity of perturbation response, it is not always sufficient to describe feature response as a single response gradient. While Gaussian perturbation can be used to sample the response curve of a single input wavelength and visualize this as a ceteris paribus-style line plot per input feature, there is much greater utility in visualizing the response curves of all input wavelengths concurrently. As such we present visualization of the response gradient per wavelength as a heatmap\figp{fig:perturb_showcase_figure}.

The response should not be assumed to be symmetric in $\xi$. Raising and lowering the spectrum by the same amount are different changes, and a non-linear retrieval, with bounded priors and degenerate parameters, can respond to them quite differently. In\fig{fig:perturb_showcase_figure}, for example, lowering the window at 1.27--1.32\,\um{} lowers the retrieved $\mathrm{H_2O}$ by much more than raising it raises $\mathrm{H_2O}$. One possible explanation is the degeneracy between $\mathrm{H_2O}$ and the cloud-top pressure\Figp{sec:metric_comparison}{Section}: a raised window could be fitted by a higher cloud deck with little change in $\mathrm{H_2O}$, whereas a grey cloud cannot lower the window below its clear-atmosphere level.

Standard feature importance visualizations, such as bar and waterfall plots, are descriptive for a few discontinuous input features, but beyond around five input features, especially with sequential data, they can become difficult and unintuitive to read and the positioning of the bars becomes overall less informative.

In order to overcome this, these heatmap plots show the change in model response, or response gradient, $\partial \vect{r}_{m,p} / \partial \xi$, where $\xi$ is the perturbation magnitude in units of $\sigma_m$ (labelled $\partial\Delta\log X/\partial\xi$, in dex\footnote{Dex (decimal exponent) is the unit of a base-10 logarithm, so a change of 1\,dex in a log abundance is a factor of 10 in the abundance, and a change of 0.5\,dex a factor of about 3.}\,$\sigma^{-1}$, in the figures). Response \textit{magnitude} (upper panel) must pass through zero at zero perturbation, since it is the difference between the zero-perturbation response and itself, so on a colour scale its information vanishes as $\xi$ approaches $0\sigma$. The response \textit{gradient} (lower panel) has no such limitation, and changes in it carry the same visual weight at every perturbation magnitude. Even large numbers of input features can therefore be condensed into a single plot that still communicates the model response.

The response gradient can be summarized for each wavelength bin $m$ and molecule $p$ by its mean over the grid, which we call the importance. With the magnitudes $\xi_k$ in ascending order and $r_{m,p,k}$ the response at $\xi_k$, this is the mean of the $K-1$ slopes between neighbouring magnitudes,
\begin{equation}
   \text{Importance}_{m,p} = \frac{1}{K-1}\sum_{k=1}^{K-1}\frac{r_{m,p,k+1} - r_{m,p,k}}{\xi_{k+1} - \xi_k},
\end{equation}
in dex\,$\sigma^{-1}$, i.e. the mean down each column of the lower panel of\fig{fig:perturb_showcase_figure}.

Although heatmaps such as these are easy to read at a glance, when it comes to analysis of these types of plots, they seem to particularly incite a certain data analysis apophenia,\footnote{The human tendency to perceive coincidental patterns in randomly distributed data, otherwise known as the clustering illusion \citep{wickham.etal2010_GraphicalInferenceInfovis}.} and it is very easy to overanalyse spurious correlation. As such, rather than rely on visual inspection, in\Fig{sec:metric}{Section} we compare the importance quantitatively against two reference metrics computed from forward models.

\subsection{Evaluation metrics}
\label{sec:metric}

\begin{figure}
    \centering
    \includegraphics[width=0.48\textwidth]{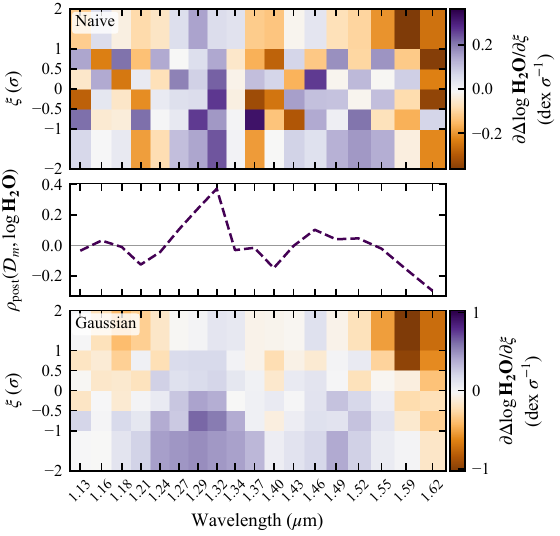}
    \caption{Upper panel shows the response gradient of the posterior mean of $\log\mathrm{H_2O}$ for HD~209458~b to \textit{naive} perturbation. Middle panel shows the local metric of $\mathrm{H_2O}$ in each wavelength bin, as in\fig{fig:Dm_logmol_H2O}. Lower panel shows the response gradient of the posterior mean of $\log\mathrm{H_2O}$ for HD~209458~b to \textit{Gaussian} perturbation, as in the lower panel of\fig{fig:perturb_showcase_figure}. Note the additional noise and inconsistency in sign of the response to the naive perturbation, compared against the much more cohesive response to the Gaussian perturbation. The two maps have separate colour scales. The Gaussian perturbation feature importance shows strong correlation with the local metric, as seen in\fig{fig:final_scatterplot} where the mean of each column of the lower panel is plotted against the local metric.}
    \label{fig:gradient_local_metric}
\end{figure}

\begin{figure}
    \centering
    \includegraphics[width=0.43\textwidth]{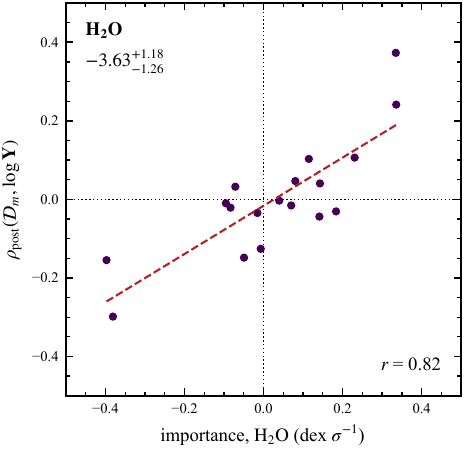}
    \caption{The importance of each wavelength bin for $\mathrm{H_2O}$, the mean response gradient under Gaussian perturbation, against the local metric of that bin. The dashed line shows the fitted linear trend, with Pearson $r = 0.82$. The label gives the retrieved $\log\mathrm{H_2O}$, the posterior median with the 16th and 84th percentiles. The full plot for all molecules can be found in the appendix in\fig{fig:final_scatterplot_all}.}
    \label{fig:final_scatterplot}

\end{figure}

Information density is not constant across any given spectral observation. While both instrument and reduction pipeline design do in some ways attempt to account for this,\footnote{Although this happens as a side effect of increasing the total amount of information we can get from an observation, and not necessarily in a way that ensures uniform information density.} there remains a large discrepancy between the information content of each wavelength bin.
Especially when we break this down to look at the abundance of a specific molecule in the retrieval, there are regions of an observation that contain almost no relevant information.

\subsubsection{Contribution functions}

\begin{figure}
    \centering
    \includegraphics[width=0.48\textwidth]{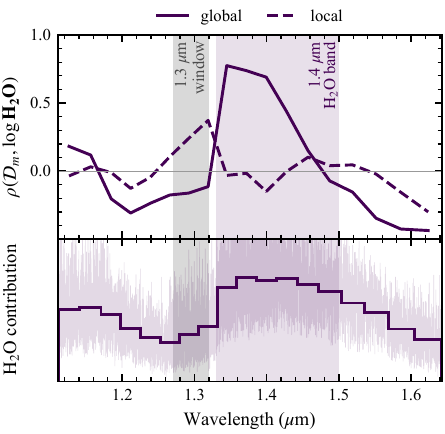}
    \caption{For $\mathrm{H_2O}$ in the HD~209458~b retrieval, the global and local metrics in each wavelength bin $m$ (upper panel), and the contribution of $\mathrm{H_2O}$ to the spectrum (lower panel). The shaded regions mark the window between the water bands at 1.27--1.32\,\um{} and the $1.4\,\um$ band of $\mathrm{H_2O}$. The global metric peaks in the band, where $\mathrm{H_2O}$ absorbs most strongly, and the local metric in the window, where the retrieved abundance of $\mathrm{H_2O}$ is decided\Figp{sec:metric_comparison}{Section}. The forward models used for each metric, and the same panels for the other gases, are given in\fig{fig:Dm_logmol}.}
    \label{fig:Dm_logmol_H2O}
\end{figure}

A forward model spectrum is composed of the individual absorption of each gas, together with collision-induced absorption (CIA) and clouds. As such, the contribution of each gas to the spectrum comes for free with the forward model. We take the contribution of a gas to be the spectrum of a forward model containing only that gas (lower panel of\fig{fig:Dm_logmol_H2O} for $\mathrm{H_2O}$, and lower panels of\fig{fig:Dm_logmol} for every gas). Contribution functions are physically motivated, and show where in the spectrum each gas absorbs, and so which wavelength bins we would expect to carry information about its abundance. They are a property of the forward model alone, independent of any observation.

\subsubsection{Global metric}
\label{sec:global_metric}
To rapidly identify the regions of importance in a way that is independent of the observed data, we define a global metric. We take a grid of forward models spanning the prior ranges of the gas abundances, with the other parameters fixed (see the captions of\fig{fig:Dm_logmol} and, for \mbox{HAT-P-26}~b,\fig{fig:HATP_correlation_contribution}), and min-max normalize each spectrum over wavelength. The global metric of gas $p$ in wavelength bin $m$ is the Pearson correlation between the normalized transit depth in that bin across all forward models, $\mathcal{D}_m$, i.e. bin $m$ of a set of forward models from the population $\mathcal{D}$ of\Fig{sec:formalising}{Section}, and their respective ground-truth abundance $\log \mat{Y}_p$,
\begin{equation}
  \text{Global}_{m,p} = \rho(\mathcal{D}_m, \log \mat{Y}_p).
\end{equation}

The normalization is necessary. With the temperature and planetary radius fixed, the overall transit depth of each forward model is set mainly by its scale height, $H = k_\mathrm{B}T/(\mu m_\mathrm{H} g)$, which every abundance changes through the mean molecular weight $\mu$ (with $k_\mathrm{B}$ the Boltzmann constant, $T$ the temperature, $m_\mathrm{H}$ the mass of a hydrogen atom and $g$ the surface gravity). Across the HD~209458~b grid the mean transit depth varies from one forward model to the next by about 50 times as much as the spectral features vary within a single spectrum. Without normalization this offset drowns out the spectral features, and the metric is almost flat in wavelength. Normalizing each spectrum to the same range removes the overall transit depth and the amplitude of the spectral features, the two quantities the scale height controls, and leaves the shape of the spectrum, which the molecular opacities set. The normalization has to be applied to each spectrum and not to each bin, since the Pearson correlation is unchanged when a bin is rescaled across forward models. This yields a metric which agrees well with the corresponding contribution functions\figp{fig:Dm_logmol}, as we would expect, since both are global measures. As a linear correlation across the prior, the global metric acts as a proxy for the information content of each wavelength bin about each molecular abundance.

\subsubsection{Local metric}
\label{sec:local_metric}
The global metric describes where each gas can shape the spectrum across the whole prior. To explain the retrieval of one observation, we instead need to know where the abundance of each gas is decided among the atmospheres that the retrieval considers plausible for that observation, i.e. its posterior. We therefore define a local metric, the same correlation computed across forward models at samples from the posterior of the retrieval for the observation $\vect{x}_\star$,
\begin{equation}
  \text{Local}_{m,p} = \rho_\mathrm{post}(\mathcal{D}^\star_m, \log \mat{Y}^\star_p),
\end{equation}
where $\mathcal{D}^\star_m$ is the transit depth in bin $m$ of the forward model at each posterior sample, $\log\mat{Y}^\star_p$ is the log abundance of gas $p$ in that sample, and $\rho_\mathrm{post}$ is the Pearson correlation with each sample weighted by its posterior weight. The samples used for each planet are given in the captions of\fig{fig:Dm_logmol} and\fig{fig:HATP_correlation_contribution}.

The two metrics can therefore disagree. Where a band is fitted equally well along a degeneracy, varying the abundance within the posterior barely changes the band, so the local metric there can be small even where the global metric peaks, as for $\mathrm{H_2O}$ in HD~209458~b\figp{fig:Dm_logmol_H2O}, discussed in\Fig{sec:metric_comparison}{Section}.

The local metric differs from the global metric in that each sample is weighted by its posterior weight. The spectra are not normalized, since every posterior sample fits the same observation and so the overall transit depth no longer varies enough to drown out the spectral features. However, like the global metric, the local metric is computed from forward models alone, so it is a local, data-driven importance\Figp{sec:data_model_driven}{Section}, whereas perturbation gives a local, model-driven one. It is important to note that this local metric provides a physically motivated reference rather than a unique ground-truth measure of feature importance, since it captures correlations within the posterior distribution rather than the retrieval's response to direct perturbations of the observed spectrum.

\subsubsection{Comparison with the importance}
\label{sec:metric_comparison}
With the importance from\Fig{sec:visualisation}{Section} and the local metric, which provides a physically motivated reference for the expected response of the retrieval for this observation, we can compare the regions expected to carry information about each molecular abundance with those to which the retrieval is most sensitive\figp{fig:final_scatterplot_all}.

When we consider $\mathrm{H_2O}$, which is the dominant molecule in this retrieval\Figp{tab:HD209_posteriors}{Table}, there is a strong agreement, $r = 0.82$, between the local metric and the recovered feature importance\figp{fig:final_scatterplot}. Both pick out the window between the water bands at 1.27--1.32\,\um{} rather than the core of the 1.4\,\um{} band, where the contribution function of $\mathrm{H_2O}$ and its global metric peak\figp{fig:Dm_logmol_H2O}. In this retrieval $\mathrm{H_2O}$ is degenerate with the cloud-top pressure, and the band core is fitted equally well along this degeneracy, so it is the window that decides the retrieved $\mathrm{H_2O}$, something the contribution functions alone do not show.

Looking to the other molecules present in the retrieval\figp{fig:final_scatterplot_all}, the recovered feature importance for $\mathrm{NH_3}$ abundance also aligns strongly with the expected importance regions ($r = 0.87$), despite its much lower retrieved abundance. For $\mathrm{CO}$ and $\mathrm{CO_2}$ almost all regions are equally unimportant in both the local metric and the Gaussian perturbation, which reflects the fact that both molecules are very challenging to constrain in the WFC3 wavelength range. For $\mathrm{CH_4}$ the local metric does highlight important bins, but the Gaussian perturbation is not able to recover these as well as with $\mathrm{H_2O}$ and $\mathrm{NH_3}$. Like $\mathrm{CO}$ and $\mathrm{CO_2}$, $\mathrm{CH_4}$ is unconstrained in this retrieval\Figp{tab:HD209_posteriors}{Table}, so its importance is weaker and noisier under any perturbation, and its few important bins also bunch towards the edges of the observation, where the kernel is truncated.

Here we are comparing a metric derived from physically constrained forward models against the Gaussian perturbation feature importance of a retrieval that is based on the same forward models. As the underlying mechanisms in both of these approaches are identical, a perfect interpretability methodology should retrieve high correlation between them. Several factors limit a hypothetically perfect recovery of feature importance, the most significant of which are the non-deterministic nature of the retrieval, the non-linearity of its response, which the importance averages over the grid of perturbation magnitudes, and the accuracy of the posterior. Given these limits, the agreement for the gases the retrieval constrains is very good, and the Gaussian perturbation technique has performed as anticipated.

\subsubsection{Comparison with naive perturbation}
\label{sec:hd209_naive}
Perturbing each wavelength bin individually (as in the naive perturbation methodology described in\Fig{sec:cetpar}{Section}), on the same grid of magnitudes, gives a much less stable response than Gaussian perturbation. While we would expect the feature response to take a low-order functional form, largely monotonic or unimodal, the naive response exhibits a very noise-dominated, multimodal gradient\figp{fig:gradient_local_metric}. Under naive perturbation the response of $\mathrm{H_2O}$ has two or more turning points in 16 of the 18 wavelength bins, whereas under Gaussian perturbation 11 of the 18 responses are monotonic or have a single turning point. The response between any one pair of neighbouring magnitudes is therefore much less reliable under naive perturbation. For $\mathrm{H_2O}$, the response-gradient maps of different pairs of magnitudes agree with each other with a mean correlation of 0.69 under Gaussian perturbation, against 0.27 under naive perturbation, and the agreement of a single pair with the local metric ranges from $r = 0.63$ to $0.81$ under Gaussian perturbation, against $0.32$ to $0.76$ under naive perturbation. A naive explanation is highly dependent on which magnitudes are sampled, which is especially important with a high sample volume or high inference latency, as for \textsc{TauREx-III}, where only a few perturbation magnitudes can be afforded.

The reason is that the retrieval cannot fit most of a single-bin perturbation. To measure this, we compare the change $\Delta x_j$ in bin $j$ of each perturbed spectrum, relative to $\vect{x}_\star$, with the change $\Delta b_j$ that it causes in the best-fitting \textsc{TauREx-III} model,
\begin{equation}
    \text{fraction of the perturbation fitted} = \frac{\sum_j \Delta b_j\,\Delta x_j/\sigma_j^2}{\sum_j \Delta x_j^2/\sigma_j^2},
    \label{eq:fraction_fitted}
\end{equation}
which is 1 when the best fit follows the perturbation exactly, i.e. when the forward model can produce the change, and 0 when the retrieval treats the change as noise. Over the 108 retrievals of each method, the best fit reproduces on average 33 per cent of a naive perturbation, against 76 per cent of a Gaussian one\Figp{sec:fraction_fitted}{Appendix}. The naive response is correspondingly closer to the sampling noise of the retrieval, with a median of 2.6 times the sampling error of the posterior mean at $|\xi| = 1\sigma$, against 7.3 times for the Gaussian perturbation, which is why its gradient changes sign between neighbouring magnitudes. Gaussian perturbation therefore gives a larger response that varies smoothly with the size of the perturbation, from modified spectra that the retrieval can fit much more effectively. For \mbox{HAT-P-26}~b\Figp{sec:hatp}{Section} the difference is even larger, with the Gaussian perturbation moving the posteriors substantially more than the naive one.

\subsection{Adding complexity with \mbox{HAT-P-26}~b}
\label{sec:hatp}
\begin{figure*}
    \centering
    \vspace{26pt}
    \includegraphics[width=0.98\textwidth]{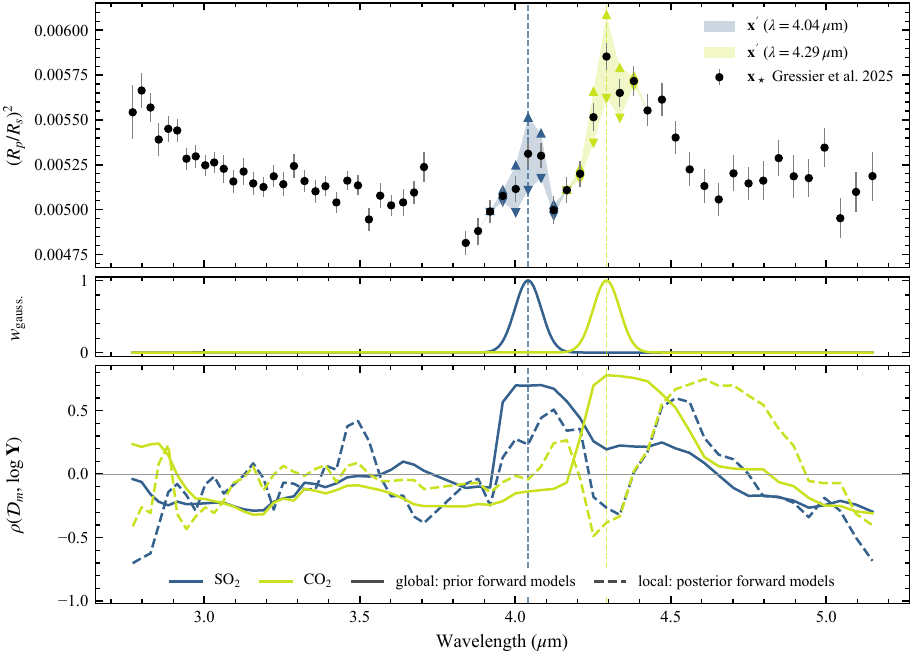}
    \caption[NIRSpec G395H observation of \mbox{HAT-P-26}~b with Gaussian perturbation]{NIRSpec G395H observation of \mbox{HAT-P-26}~b from \cite{gressier.etal2025_JWSTTSTDREAMSSulfur} with Gaussian perturbation added. We specifically target the $4.04\,\um$ and $4.29\,\um$ wavelength bands as these represent narrow $\mathrm{SO_2}$ and $\mathrm{CO_2}$ spectral features respectively. The top panel shows the \mbox{HAT-P-26}~b reduction and the modified data points. The middle panel shows the Gaussian kernel, with the same FWHM of $0.1\,\um$ as for HD~209458~b and a magnitude of $\pm3\sigma$. The lower panel shows the global metric (solid) and the local metric (dashed) of $\mathrm{SO_2}$ and $\mathrm{CO_2}$, calculated as for HD~209458~b in\fig{fig:Dm_logmol} from forward models drawn from the \mbox{HAT-P-26}~b priors and from the \textsc{TauREx-III} posterior of this observation respectively. Both metrics of every gas are shown in\fig{fig:HATP_correlation_contribution}.}
    \label{fig:observation}
\end{figure*}

Having explored interpretability with a simple retrieval that only contains one strong detection at low resolution, we now explore a more complex case, with several molecules in detection and a higher-resolution observation. First we repeat the experiment from\fig{fig:Dm_logmol}, by generating a forward model grid based on the \mbox{HAT-P-26}~b priors\Figp{tab:hatp26_priors}{Table}, and then calculate the correlation between a given wavelength bin across all forward models and their respective ground-truth abundance at a global and local scale. Again, as in\Fig{sec:local_metric}{Section}, the local metric uses the \textsc{TauREx-III} posterior from this observation. The full results of this can be seen in the appendix in\fig{fig:HATP_correlation_contribution}.

We choose to focus on two molecules here, $\mathrm{SO_2}$ and $\mathrm{CO_2}$. Both show strong detection in the retrieval\Figp{tab:HATP_posteriors}{Table} and have clearly defined spectral features at the $4.04\,\um$ and $4.29\,\um$ wavelength bands respectively. Selecting a subset of the full interpretability study allows us to examine the posterior degeneracies in more detail than was considered in the HD~209458~b methodology. As in\Fig{sec:metric_comparison}{Section}, we take the local metric as the expected response of the retrieval, and at each band we test whether the response has the sign it predicts.\fig{fig:observation} shows the \cite{gressier.etal2025_JWSTTSTDREAMSSulfur} base \mbox{HAT-P-26}~b observation, examples of perturbed spectra with perturbation magnitude $\xi = \pm3\sigma$, and the global and local metrics of both gases. At $4.04\,\um$ the local metric is positive for $\mathrm{SO_2}$ and close to zero for $\mathrm{CO_2}$, so perturbation centred on this band should move the retrieved $\mathrm{SO_2}$ in the same direction as the perturbation and leave $\mathrm{CO_2}$ unchanged. The global metric agrees, as this is the band in which $\mathrm{SO_2}$ absorbs. At $4.29\,\um$ the local metric is negative for both $\mathrm{CO_2}$ and $\mathrm{SO_2}$, so raising the transit depth around this band should lower the retrieved abundances of both gases, even though this is the band in which $\mathrm{CO_2}$ absorbs most strongly and the global metric of both gases is positive. This makes $4.29\,\um$ a strong test of the method. We retain the same FWHM of $0.1\,\um$ as in\Fig{sec:hd209_methodology}{Section}. To test these predictions, we visualize the full marginal posteriors of $\mathrm{SO_2}$ and $\mathrm{CO_2}$ so that we can analyse any correlation that may otherwise not be captured by the posterior mean alone.

We also retrieve the naively perturbed spectra at the same magnitudes for the same wavelength bins, so that we can draw comparisons between the two methods.

\begin{figure*}
    \vspace{16pt}
    \centering
        \includegraphics[width=\textwidth]{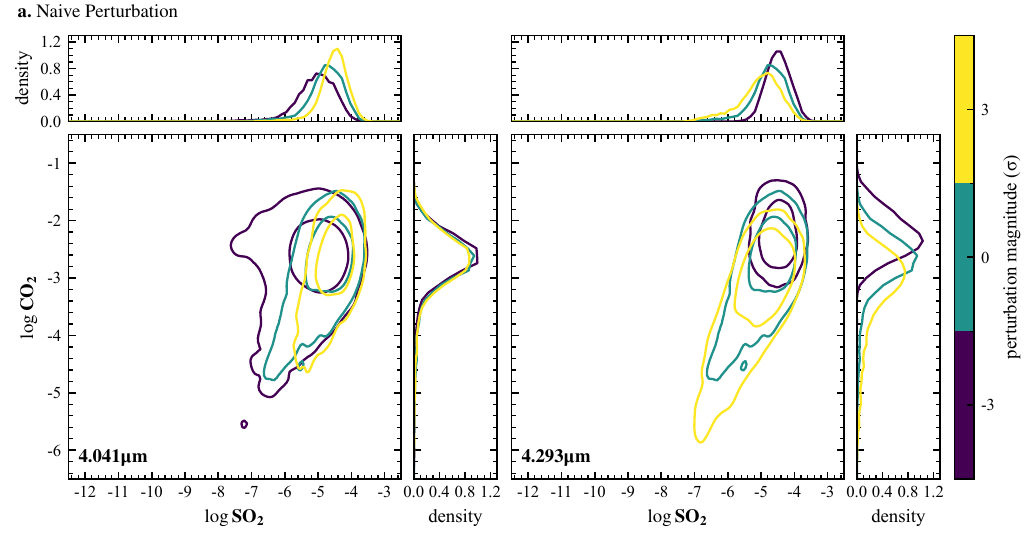}

        \includegraphics[width=\textwidth]{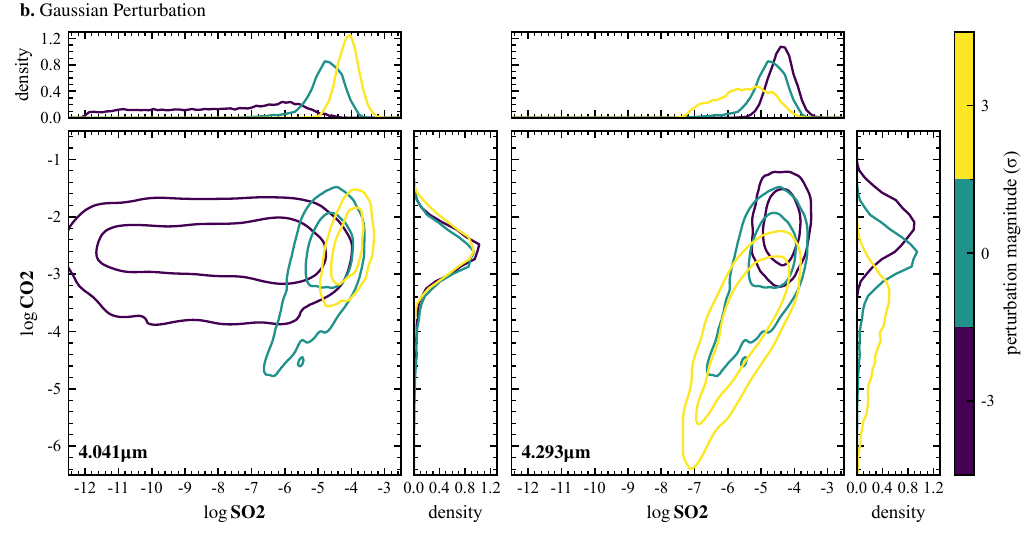}
        \caption{Posterior contour plots for the retrieved abundance of $\mathrm{CO_2}$ and $\mathrm{SO_2}$ under varying perturbation magnitudes.
        The upper two panels \textbf{a.} show the naive perturbation and the lower two panels \textbf{b.} show the Gaussian perturbation. The left two panels show the effects of the perturbation centred on the $4.04\,\um$ band, which targets the $\mathrm{SO_2}$ feature, and the right two panels show the effects of the perturbation centred on the $4.29\,\um$ band, which targets the $\mathrm{CO_2}$ feature. The teal contours show a perturbation magnitude of $\xi = 0\sigma$, i.e. the unperturbed baseline retrieval, and are identical in all panels.}
        \label{fig:marginal_posteriors}
\end{figure*}

\subsubsection{Naive and Gaussian responses}
As for HD~209458~b\Figp{sec:hd209_naive}{Section}, the posteriors of\fig{fig:marginal_posteriors} show a much larger response to Gaussian perturbation than to naive perturbation. Both the changes to the posterior mean and those to the uncertainty of the posteriors are substantially larger. The most striking example of this is the $\mathrm{SO_2}$ posterior under $-3\sigma$ Gaussian perturbation to the $4.04\,\um$ wavelength bin, as this increases the uncertainty in $\log \mathrm{SO_2}$ by a factor of two over the naive perturbation, taking the posterior mean with it. The same effect is noted in the other direction, in the $+3\sigma$ Gaussian perturbation, which shrinks the uncertainty relative to the naive perturbation. In the $4.29\,\um$ wavelength bin, this effect is also present in the retrieved $\mathrm{CO_2}$, and can be seen most clearly from the histograms.

These changes to the value of the posterior mean, driven in part by changes to the uncertainty, may be the reason why in\fig{fig:marginal_posteriors} we see a greater response gradient with negative perturbation values than with positive, as the effects of these two phenomena compound.

\subsubsection{The \texorpdfstring{$\mathrm{SO_2}$}{SO2} band at \texorpdfstring{$4.04\,\um$}{4.04 \textmu m}}
Regarding the hypothesis for this experiment, we confirm that by introducing perturbation centred on the $4.04\,\um$ wavelength band we are able to see a corresponding change in the retrieved posterior for $\mathrm{SO_2}$ in both the naive and the Gaussian case. The direction of the perturbation response is positively correlated with the perturbation magnitude, as both metrics predict. As we have already mentioned, in the Gaussian case this effect is far more pronounced, but despite this, there is no notable change in the posterior of $\mathrm{CO_2}$ even under Gaussian perturbation. Again, this is what both metrics predict.

\subsubsection{The \texorpdfstring{$\mathrm{CO_2}$}{CO2} band at \texorpdfstring{$4.29\,\um$}{4.29 \textmu m}}
At $4.29\,\um$ the response also has the sign the local metric predicts. In both the naive and the Gaussian case the retrieved posteriors of $\mathrm{CO_2}$ and $\mathrm{SO_2}$ change, and the direction of the response is negatively correlated with the perturbation magnitude. Increasing the $(R_\mathrm{p}/R_\mathrm{s})^2$ of the observation centred on $4.29\,\um$ decreases the retrieved abundance of both gases, and of $\mathrm{H_2O}$ as well, as its local metric also predicts\figp{fig:HATP_correlation_contribution}. Unlike $\mathrm{CO_2}$ under perturbation at $4.04\,\um$, $\mathrm{SO_2}$ responds clearly to the perturbation of the other gas's band, a first sign that the response at $4.29\,\um$ is not set by the opacity of $\mathrm{CO_2}$ alone. To unpack it,\fig{fig:4.293_HATP_full_posterior_gauss} shows the full posterior for the perturbation at $4.29\,\um$.

\begin{figure}
    \centering
    \includegraphics[width=0.48\textwidth]{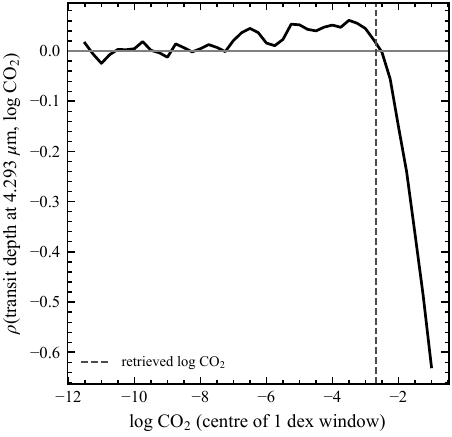}
    \caption{The correlation between the transit depth in the $4.29\,\um$ bin, the core of the $\mathrm{CO_2}$ band, and $\log\mathrm{CO_2}$, as in the global metric\figp{fig:Dm_logmol} but on the raw spectra and within sliding 1-dex windows of $\log\mathrm{CO_2}$, across the 100\,000 forward models of the training set of\Fig{sec:ml}{Section}, whose abundances are drawn from the retrieval priors. The dashed line marks the retrieved $\log\mathrm{CO_2}$, the posterior median of $-2.68$.}
    \label{fig:co2_core_window_metric}
\end{figure}

\begin{figure*}
    \centering
    \includegraphics[width=0.96\textwidth]{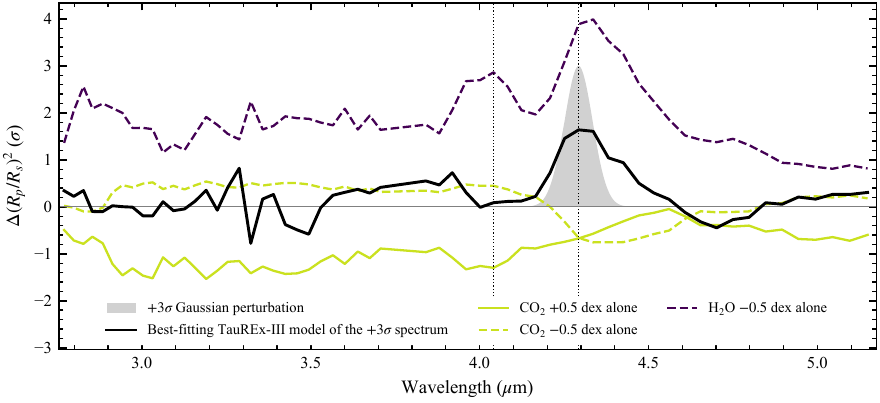}
    \caption{Change in the spectrum, in units of each bin's uncertainty, relative to the best-fitting \textsc{TauREx-III} model of the unperturbed spectrum (its maximum a posteriori sample). The shaded curve is the $+3\sigma$ Gaussian perturbation at $4.29\,\um$. The coloured lines each change one abundance by 0.5\,dex with all else fixed, raising (solid) or lowering (dashed) $\mathrm{CO_2}$ and lowering $\mathrm{H_2O}$. Raising and lowering $\mathrm{CO_2}$ alone both lower the core of the $\mathrm{CO_2}$ band at $4.29\,\um$, whereas lowering $\mathrm{H_2O}$ raises it, as the smaller mean molecular weight stretches the atmosphere. The black line is the best-fitting \textsc{TauREx-III} model of the perturbed spectrum, which raises the band core by about half of the perturbation while leaving the rest of the spectrum almost unchanged. Dotted vertical lines mark the $\mathrm{SO_2}$ and $\mathrm{CO_2}$ bands at 4.04 and $4.29\,\um$.}
    \label{fig:co2_band_spectral_changes}
\end{figure*}

Under the $+3\sigma$ perturbation at $4.29\,\um$ the retrieval does not add $\mathrm{CO_2}$. Instead, the posterior medians of $\mathrm{H_2O}$, $\mathrm{CO_2}$ and $\mathrm{SO_2}$ fall by 1.21, 1.12 and 0.76\,dex respectively, so that the mean molecular weight $\mu$ falls by 0.6, and the temperature rises by 45\,K, while the planetary radius decreases by only $0.008\,R_\mathrm{J}$\figp{fig:4.293_HATP_full_posterior_gauss}. Each of these changes increases the scale height $H$, and with it the height of every spectral feature above the continuum, so the retrieval raises the band by stretching the atmosphere. It cannot raise it by adding $\mathrm{CO_2}$. Extra $\mathrm{CO_2}$ adds opacity, but it also increases $\mu$ and so shrinks the scale height, and the core of this strong band lies many scale heights above the pressure at which the planetary radius is defined, which makes it especially sensitive to the scale height. Across forward models drawn from the retrieval priors, the transit depth of the band core increases with $\mathrm{CO_2}$, on average, up to $\log\mathrm{CO_2} \approx -2.5$, and decreases above it\figp{fig:co2_core_window_metric}. \mbox{HAT-P-26}~b, with a posterior median of $\log\mathrm{CO_2} = -2.68$, lies at this turnover.\fig{fig:co2_band_spectral_changes} shows that changing $\mathrm{CO_2}$ alone in either direction lowers the band core, while lowering $\mathrm{H_2O}$, and with it $\mu$, raises it. The best-fitting model of the perturbed spectrum raises the band core by about half of the perturbation while leaving the rest of the spectrum almost unchanged. The fall in $\mathrm{CO_2}$ and $\mathrm{SO_2}$ then stops the rest of the $\mathrm{CO_2}$ band and the $\mathrm{SO_2}$ feature, which would otherwise rise with the scale height, from moving away from the data. The $-3\sigma$ perturbation does the reverse, raising the posterior medians of $\mathrm{H_2O}$, $\mathrm{CO_2}$ and $\mathrm{SO_2}$ by 0.41, 0.52 and 0.34\,dex respectively, and $\mu$ by 1.0, to compress the atmosphere.

This is a form of the well-known degeneracy between temperature, mean molecular weight and the amplitude of spectral features through the scale height \citep{benneke.seager2012_ATMOSPHERICRETRIEVALSUPEREARTHS, wit.seager2013_ConstrainingExoplanetMass}. The strong correlation between $\mathrm{CO_2}$ and $\mathrm{H_2O}$ in the posterior\figp{fig:4.293_HATP_full_posterior_gauss} has the same origin, since both set $\mu$, and so does the anticorrelation between $T$ and $R_\mathrm{p}$, since a warmer, more extended atmosphere can be compensated by a smaller reference radius \citep{griffith2014_DisentanglingDegenerateSolutions}.

In summary, the Gaussian perturbation recovered the expected response of the retrieval at both bands, in agreement with the local metric, and at $4.29\,\um$ it revealed how the retrieval responds to the core of the $\mathrm{CO_2}$ band through the scale height rather than through the opacity of $\mathrm{CO_2}$, by way of degeneracy.

\section{Application to machine learning}
\label{sec:ml}
\begin{figure*}
    \centering
    \includegraphics[width=\textwidth]{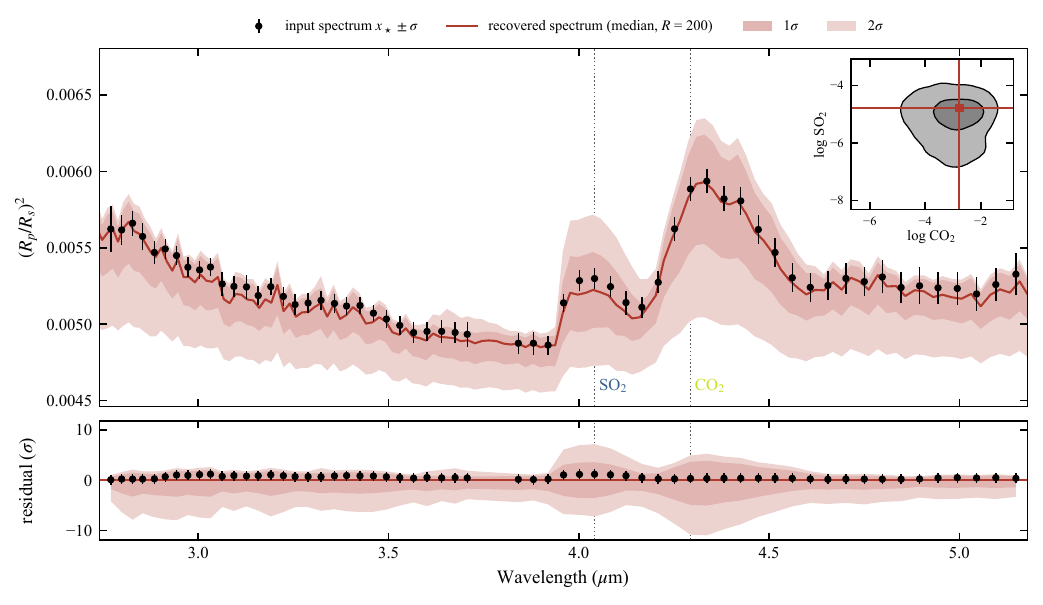}
    \caption[Machine learning retrieval of a simulated observation of \mbox{HAT-P-26}~b at NIRSpec G395H resolution.]{Machine learning retrieval of a simulated observation of \mbox{HAT-P-26}~b at NIRSpec G395H resolution. The posterior combines Monte Carlo dropout with observational noise: 20\,000 noisy realizations of the simulated spectrum ($\sigma_m$ per bin) are each passed through the network once with an independent dropout mask, and \textsc{TauREx-III} spectra at $R=200$ are generated from 5\,000 of these predictions. The main panel shows the median and the $1\sigma$ and $2\sigma$ ranges of these spectra, and the lower panel shows the residuals relative to the median in units of $\sigma_m$. The inset shows the $1\sigma$ and $2\sigma$ contours of the joint posterior of $\mathrm{SO_2}$ and $\mathrm{CO_2}$. Dotted vertical lines indicate the locations of spectral features for $\mathrm{SO_2}$ and $\mathrm{CO_2}$. In the inset, the red lines indicate the input values of the simulated spectrum, the \textsc{TauREx-III} posterior means in\Fig{tab:HATP_posteriors}{Table}. The full posterior corner plot can be seen in\fig{fig:HATP_full_posterior_ML}.}
    \label{fig:hatp_reconstructed_spec}
\end{figure*}

\begin{figure}
    \centering
    \includegraphics[width=0.48\textwidth]{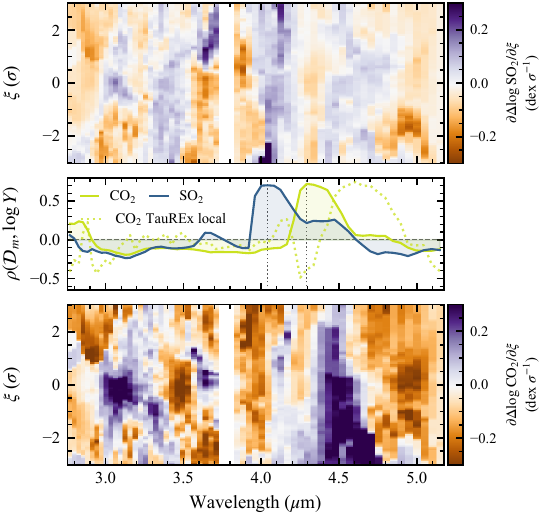}
    \caption{Response-gradient maps for $\mathrm{SO_2}$ and $\mathrm{CO_2}$ shown against the global metric of both gases, computed as in\fig{fig:Dm_logmol} over the training spectra of the network. The dotted line is the local metric of $\mathrm{CO_2}$ from the \textsc{TauREx-III} posterior of the observed spectrum\Figp{sec:local_metric}{Section}, shown only as a visual guide to where the abundance of $\mathrm{CO_2}$ is decided for this atmosphere. It is not a local metric of the network, which has no calibrated posterior\Figp{sec:ml_simulated}{Section}. The network's prediction is the mean over a fixed set of 1\,000 Monte Carlo dropout masks. The colour scale is consistent between the upper and lower plots.
}
    \label{fig:hatp_corner_co2_so2}
\end{figure}

To evaluate this perturbation method on a genuine machine learning retrieval, we train a deliberately simple one-dimensional CNN based on the baseline released for the ESA-Ariel Data Challenge 2022 (ADC; \citealt{yip.etal2022_ESAArielDataChallenge}). Details of the model architecture and training are available in\Fig{sec:cnn}{Appendix}.

\subsection{Retrieval of the simulated observation}
\label{sec:ml_simulated}
We use the network to predict on the simulated observation of\Fig{sec:dim_reduction}{Section}, the forward model at the \textsc{TauREx-III} posterior means, since the real spectrum contains instrumental and reduction effects that are absent from the training set. Like the stored training spectra, this simulated observation is noiseless.

To obtain a posterior we keep the baseline's treatment of uncertainty: dropout stays active at inference \citep[Monte Carlo dropout;][]{gal.ghahramani2016_DropoutBayesianApproximation}, so that each pass through the network drops a different random 10 per cent of the dense units and the spread of the passes is taken as the network's predictive uncertainty. Where the ADC baseline passes one observed spectrum through the network 5\,000 times, we draw 20\,000 realizations of the simulated observation with Gaussian noise of standard deviation $\sigma_m$, as in training, and pass each through the network once with its own dropout mask, so that the posterior combines the network's uncertainty with the observational noise. Monte Carlo dropout gives only an approximate posterior, and one that is known to be overconfident. Here the spread of the dropout passes does not widen for gases the spectrum cannot constrain, so the intervals are too narrow and the posterior is more certain of its predictions than the data allow. Averaged over the ten gases, the central 68 per cent interval contains the true abundance for only 28 per cent of test spectra (and the 95 per cent interval for 49 per cent). We therefore use these intervals to show how the network's prediction varies, not as calibrated error bars.

We then use \textsc{TauREx-III} forward model to reconstruct the predicted spectrum based on the abundance posteriors from the machine learning model. The reconstructed spectrum plotted against the simulated observation points can be seen in\fig{fig:hatp_reconstructed_spec} alongside a subset of the joint posterior of this machine learning retrieval, with the full posterior in\fig{fig:HATP_full_posterior_ML}. The network recovers $\mathrm{H_2O}$, $\mathrm{CO_2}$ and $\mathrm{SO_2}$ within $1\sigma$ of the input values (posterior medians with the 16th and 84th percentiles of $-1.86^{+0.54}_{-1.25}$, $-2.97^{+0.74}_{-0.94}$ and $-5.04^{+0.49}_{-0.81}$, against $-1.63$, $-2.77$ and $-4.78$), but not $\mathrm{H_2S}$ ($-8.26^{+0.53}_{-0.40}$ against $-4.91$), which \textsc{TauREx-III} itself constrains only weakly (a posterior median of $-3.65^{+0.60}_{-4.36}$). These intervals are dominated by the Monte Carlo dropout spread rather than by the observational noise. The median reconstructed spectrum lies slightly below the simulated observation in every bin, by at most $1.2\sigma$, mostly near $3\,\um$ and in the $\mathrm{SO_2}$ band at 3.96--4.12\,\um{}, so overall the network retrieves a convincing fit to the data. The posterior extends further below the data than above it. The lowest spectra come from draws with high abundances of $\mathrm{H_2O}$, $\mathrm{CO_2}$ and $\mathrm{SO_2}$, which raise the mean molecular weight and so, at fixed temperature and planetary radius, shrink the scale height and lower the whole spectrum.

\subsection{Perturbing the network}
For the network we recompute the global metric of\Fig{sec:global_metric}{Section} over its training spectra, rather than the 486 forward models of\fig{fig:HATP_correlation_contribution}, so that it describes the spectra the network has learned from. For $\mathrm{SO_2}$ and $\mathrm{CO_2}$ the two versions are almost identical.

The perturbation methods do not require $f$ to be deterministic, only that a perturbation changes its output by more than the model's own randomness does. With a fresh dropout mask at every evaluation a single pass through the network has a standard deviation of about 0.7\,dex in $\log\mathrm{SO_2}$, more than the full response to a $\pm3\sigma$ perturbation of the $4.04\,\um$ band. Averaging 1\,000 fresh masks per evaluation is enough to recover the importance of every bin, a mean over the whole $\pm3\sigma$ grid, but not the response-gradient maps, which resolve the much smaller change between adjacent $0.06\sigma$ steps\figp{fig:fresh_dropout}. For all results in the main text we therefore define $f(\vect{x})$ as the mean prediction over a fixed set of 1\,000 dropout masks, drawn once and reused for every input, so that the dropout noise cancels when two predictions are compared and any change in $f$ is caused by the change in the spectrum alone. This gives clean response-gradient maps, and the same importance as fresh masks. Where the randomness of a black-box cannot be fixed in this way, the perturbations must instead be large enough, or each evaluation averaged enough, for the response to stand out from it.

We apply the Gaussian perturbation methodology to the trained network for the simulated \mbox{HAT-P-26}~b observation, and reconstruct its response-gradient maps.\fig{fig:hatp_corner_co2_so2} shows the response-gradient maps for our target molecules $\mathrm{SO_2}$ and $\mathrm{CO_2}$ against this global metric, together with the local metric of $\mathrm{CO_2}$ from the \textsc{TauREx-III} posterior of the observed spectrum. The maps are almost unchanged if dropout is instead switched off at inference, so they do not depend on this choice. The $\mathrm{SO_2}$ map shows strong importance centred on $4.04\,\um$, rejection of the $\mathrm{CO_2}$ feature at $4.29\,\um$, as well as some tenuous agreement with the secondary $\mathrm{SO_2}$ feature at $3.65\,\um$ under high positive perturbation, all of this in accordance with what we would expect. The $\mathrm{CO_2}$ map, however, shows strong importance centred on the upper tail of the expected $\mathrm{CO_2}$ feature, and almost none at the core of the band at $4.29\,\um$, where it is slightly negative. The \textsc{TauREx-III} local metric of $\mathrm{CO_2}$ for the observed spectrum shows the same pattern, negative at the band core and positive on its long-wavelength side, where it peaks near $4.6\,\um$, as expected from the scale-height response of\Fig{sec:hatp}{Section}. The $\mathrm{CO_2}$ map also shows some rejection of the $4.04\,\um$ $\mathrm{SO_2}$ feature, and is overall noisier than the $\mathrm{SO_2}$ map, which is to be expected, given that more gases compete for the bins around $4.29\,\um$. In the global metric\figp{fig:hatp_corner_co2_so2}, $\mathrm{SO_2}$ also correlates with these bins, as noted for \textsc{TauREx-III} in\Fig{sec:hatp}{Section}.

We are unable to compute the data-driven local metric for the CNN retrieval, as the Monte Carlo dropout employed by the machine learning framework does not produce calibrated posteriors. The posterior intervals are too narrow to describe the local posterior region well enough to produce an accurate metric. The local metric of $\mathrm{CO_2}$ shown in\fig{fig:hatp_corner_co2_so2} is that of the \textsc{TauREx-III} posterior of the observed spectrum, a reference for this atmosphere rather than for the network. The network's response to the band core can be compared with that of \textsc{TauREx-III} in\Fig{sec:hatp}{Section}, but only qualitatively, as the two retrievals do not see the same spectrum. \textsc{TauREx-III} retrieves the observed spectrum, whereas the network is given the simulated observation of\Fig{sec:ml_simulated}{Section}. This forward model is close to the observation but not identical to it, and the temperature, planetary radius and cloud-top pressure, which \textsc{TauREx-III} retrieves, are fixed in the network's training set. This being said, a $+3\sigma$ Gaussian perturbation centred on $4.29\,\um$ lowers the predicted $\log\mathrm{H_2O}$, and a $-3\sigma$ perturbation raises it, in the same direction as the \textsc{TauREx-III} posterior of $\mathrm{H_2O}$. The predicted $\mathrm{SO_2}$ also moves in the same direction, and both it and the predicted $\mathrm{CO_2}$ change much less than $\mathrm{H_2O}$. Since the network has never seen the temperature vary, it cannot raise the scale height through the temperature as \textsc{TauREx-III} does. It can still do so through the mean molecular weight, which the abundances set, and it does so mainly through $\mathrm{H_2O}$, the gas that dominates $\mu$\figp{fig:co2_band_spectral_changes}. The network has therefore learned the same scale-height response from its training spectra, which supports the conclusion that this response is a property of the forward models rather than of the retrieval method.

\subsection{Computational cost}
The benefits of using this technique with machine learning retrieval frameworks (or any other framework able to process spectra with low latency) come from the resolution of data that can be extracted. In this case, we are able to compute the response profile at a resolution of $0.06\sigma$, which is of the order of $1\times10^{-5}\,(R_\mathrm{p}/R_\mathrm{s})^2$, in under 3 seconds on the CPU of a \texttt{Mac M4 Pro 48\,GB}. Since dropout acts only on the dense layers, each spectrum passes through the convolutional layers and the first dense layer once, and only the last two dense layers are evaluated for each of the 1 000 masks. This gives exactly the predictions of 1 000 separate passes for about 1/30 of the computation. This is without GPU acceleration: the network was trained on the GPU (using \textsc{mps}), but we ran inference on the CPU so that the comparison with \textsc{TauREx-III}, which cannot use the GPU, is like for like; on the GPU the same analysis takes about 1 second. The grid has 101 perturbation magnitudes, one of which ($\xi=0\sigma$) is shared by every bin, so this is $100$ perturbation steps $\times$ $60$ wavelength bins $+\,1$ unperturbed retrieval $= 6\,001$ retrievals in total, each the mean of 1\,000 Monte Carlo dropout passes ($6\times10^{6}$ passes through the network). Performing this analysis at the same perturbation resolution using \textsc{TauREx-III}, in the same configuration as in\Fig{sec:hatp}{Section}, would take about $3.9\times10^4$ hours.\footnote{Based on a mean retrieval time of 6.5 hours, measured over the 8 perturbed retrievals of\Fig{sec:hatp}{Section}, again using a \texttt{Mac M4 Pro 48\,GB} with \textsc{mpi} acceleration.}

\section{Limitations on retrievals and interpretability}

This work raises the question of whether future black-box retrievals should be held up against the same degeneracies that we currently struggle with, or given scope to explore different solutions for breaking them. These degeneracies are largely physical rather than specific to current forward model calculations, so it seems pertinent that future black-box models be judged on handling them in the same way as current methods. Indeed, in\Fig{sec:ml}{Section} we find that a neural network trained on the same forward models has learned the same scale-height response as \textsc{TauREx-III}.

If we are to take \textsc{TauREx-III} as a gold standard, how should an interpretability method be judged? The forward model gives us physical expectations of where each gas can leave a signature, through the contribution functions and the global metric, but an explanation of one spectrum should be judged against the response of the retrieval to that spectrum. For a Bayesian retrieval the posterior provides this reference, the local metric, and the Gaussian perturbation recovers it, degeneracy included. In this work we seek only to validate the interpretability method itself, and this agreement is what allows us to apply it to retrievals for which no such reference exists.

The local metric is not, however, a substitute for perturbation. It needs a posterior that describes the spectrum, which a machine learning retrieval may not provide\Figp{sec:ml}{Section}, and a forward model at each posterior sample, so when the retrieval is cheaper to evaluate than the forward model, as a neural network is, perturbation is also the faster of the two. The two also describe different things, data- and model-driven importance\Figp{sec:data_model_driven}{Section}, and they agree here because both \textsc{TauREx-III} and the machine learning networks are accurate retrievals (over these specific observations). Therefore, we need to build trust in these explanation methods should we desire to use them going forward where validation with other metrics is not possible.

\section{Conclusions}
This work begins to unpack some of the issues with existing interpretability frameworks for retrievals, but is by no means a comprehensive rebuttal of the techniques, nor does it propose an exhaustive successor. Nevertheless, as both the volume and resolution of data increase, and as we begin to ask more complex questions about distant worlds, some combination of machine learning and Bayesian retrieval frameworks seems likely, if not inevitable, in the future of retrievals. As such, we have a long way to go in the exploration of suitable interpretability metrics to describe our confidence in the `reasoning' of these methods. Beginning to explore ways by which we can mitigate or circumvent the complex effects of the correlation which is so prevalent within spectral data marks a crucial step in the direction towards transparency in all black-box retrieval frameworks.

The experiments outlined here and the Gaussian perturbation technique seek to contribute to this. We show evidence for the correlated nature of spectral observations, and characterize this as approximately Gaussian in relation to wavelength separation. We demonstrate the susceptibility of ceteris paribus profiles to correlation of this nature; their propensity to generate out-of-distribution samples stems from perturbing each input feature independently, which does not account for inherent \textit{intra}sample correlation. For HD~209458~b, \textsc{TauREx-III} fits single-bin perturbations poorly, so its response to naive perturbation is weak, dominated by noise and highly dependent on the perturbation magnitudes selected. We then propose Gaussian perturbation, which perturbs neighbouring bins together, following this approximately Gaussian correlation. \textsc{TauREx-III} fits the modified spectra from this technique well, capturing on average 76 per cent of the perturbation in its best-fitting models, compared with 33 per cent for the naive methodology, and Gaussian perturbation is far more robust, giving a larger response that varies smoothly with the size of the perturbation and is largely independent of the perturbation magnitudes chosen. We demonstrate the application of these techniques using \textsc{TauREx-III} as a black-box retrieval. The importance recovered by Gaussian perturbation agrees with the local metric, the response expected of the retrieval for each observation, and shows how the retrieval uses each spectrum, which the contribution functions of its forward model alone do not. For HD~209458~b the retrieved $\mathrm{H_2O}$ is decided by the window between the water bands, through its degeneracy with the cloud-top pressure, and for \mbox{HAT-P-26}~b the response to the $4.29\,\um$ $\mathrm{CO_2}$ band is set by the degeneracy between temperature, mean molecular weight and feature amplitude. This is the sense in which even a retrieval built on a physical forward model is a black-box, and the method recovers its local response wherever that response can be checked. In addition to this, we present the resulting interpretability as a novel visualization which we hope is more intuitive than standard feature importance visualizations in conveying the information, given the large number of input features.

To demonstrate the efficacy of the method explicitly on machine learning retrievals, we apply it to a CNN trained on \textsc{TauREx-III} forward models of \mbox{HAT-P-26}~b, with Monte Carlo dropout uncertainties. The response maps use its mean prediction over a fixed set of 1\,000 dropout masks, which removes the dropout noise from the comparisons; averaging over fresh masks gives the same importance, so the method does not depend on the network being deterministic. The full response-gradient maps take under 3 seconds to compute on a laptop CPU, so the technique is very lightweight given a trained model. The network's response-gradient maps localize the $\mathrm{SO_2}$ feature at $4.04\,\um$, and its response to the core of the $4.29\,\um$ $\mathrm{CO_2}$ band has the same direction as that of \textsc{TauREx-III}, through the predicted $\mathrm{H_2O}$ abundance, even though the temperature is fixed in its training set. Its $\mathrm{CO_2}$ importance sits on the long-wavelength side of the $4.29\,\um$ band rather than at its core, where the \textsc{TauREx-III} local metric of the observation also places it.

To conclude, we hope that these experiments will help to lay the groundwork for further exploration into how we can collectively improve confidence and cohesion in our retrieval strategies going forward, in support of the population-level studies of the future.

\section*{Acknowledgements}

With thanks to Sushuang Ma for work in identifying the planet candidates, Jack Davey for help with everything error-bar-related, and Sergey Yurchenko and everyone at the ExoMol team for preparing the line lists used in generating molecular contributions and forward models.

\section*{Data Availability}


\textsc{Python} implementation of the Gaussian perturbation methods as they are described in this paper can be found at
\url{https://github.com/Jools-Clarke/perturb}. Access to any further data, such as data from the figures in tabular format, can be provided upon request.

\section*{Conflict of Interest}

The authors recognize affiliations with the research group maintaining the retrieval framework \textsc{TauREx-III}, and as such the decision to use this specific framework in this work was not made impartially. We encourage other work which tests this methodology with alternative retrieval frameworks. There are no other conflicts of interest.

\bsp	
\label{lastpage}
\bibliographystyle{rasti}
\bibliography{references} 

\vfill
\eject

\pagebreak
\clearpage



\appendix
\clearpage
\section{Brief overview of exoplanet observation strategy}
\label{sec:retrieval_background}
The first transit observations were performed by ground-based wide-field telescopes \citep{charbonneau.etal1999_DetectionPlanetaryTransits}; because they were only looking for dimming effects in white light, they could survey multiple stars simultaneously over long time periods. In the modern era transit observations are predominantly taken using space-based instruments which target a single stellar host with a spectrograph when the planet is in transit, according to a pre-calculated ephemeris. As such the raw data collected by transit observations are \textit{flux} per \textit{wavelength} over \textit{time}. The time component of these observations can be removed by a data reduction \citep{bell.etal2022_EurekaEndtoEndPipeline} to produce a fixed spectrum, which represents the idealized mid-transit where the planet and star are in exact syzygy with the observer and appear as concentric discs. As such the units with which we describe the transit depth in these observations are $(R_\mathrm{p}/R_\mathrm{s})^2$, where $R_\mathrm{p}$ and $R_\mathrm{s}$ are the planetary and stellar radii. This observation is commonly modelled as a one-dimensional snapshot of a globally homogeneous atmosphere at the terminator (sunrise on one half of the planet's circumference and sunset on the other), although differences between the morning and evening limbs can bias such retrievals \citep{macdonald.etal2020_WhyItSo, espinoza.jones2021_ConstrainingMorningsEvenings}. From this, the chemical composition of the atmosphere can be estimated, based on the spectral absorption properties of different molecular species which can be calculated through a combination of laboratory experiments and ab initio calculations \citep{tennyson.yurchenko2018_ExoMolAtlasMolecular}.

Both the volume and quality of transit data recorded have been rapidly increasing, and with the advent of \textit{JWST} \citep{gardner.etal2023_JamesWebbSpace} and now the upcoming \textit{Ariel} mission \citep{tinetti.etal2021_ArielEnablingPlanetary}, this trend stands to continue \citep{lustig-yaeger.etal2023_JWSTTransmissionSpectrum}. These missions, and others like them over the coming years, will deliver unprecedented quantity and quality of spectral data of this nature, allowing for characterization of a significantly larger percentage of the planets detected to date \citep{edwards.tinetti2022_ArielTargetList}, but also characterization in much more detail of those which have already been studied \citep{tsiaras.etal2019_WaterVapourAtmosphere, benneke.etal2019_WaterVaporClouds, madhusudhan.etal2023_CarbonbearingMoleculesPossible}.\footnote{If we take \mbox{K2-18}~b as an example, in \citet{tsiaras.etal2019_WaterVapourAtmosphere} a total of 17 data points are recorded from \textit{HST} WFC3; later, in \citet{benneke.etal2019_WaterVaporClouds}, it is 20 data points in total (17 from \textit{HST} WFC3, 1 from \textit{Kepler} and 2 from the \textit{Spitzer} Infrared Array Camera), yet with the launch of \textit{JWST} we see \citet{madhusudhan.etal2023_CarbonbearingMoleculesPossible}, where 4\,411 data points are recorded, 1\,010 from the \textit{JWST} Near Infrared Imager and Slitless Spectrograph and 3\,401 from \textit{JWST} NIRSpec.} This creates both opportunity and challenge for the field: the wealth of new data will facilitate better understanding at a population level, allowing us to study planets in the context of a wider sample. However, analysis methods must scale efficiently in order to manage this magnitude of data.

The current gold standard for interpreting exoplanet observations is iterative optimization of the inverse problem. It is much less complex to simulate a transit observation given parameters about the system than it is to infer these system parameters from an observation. A forward model is a physics-driven atmospheric model capable of taking a set of planetary and system parameters and producing a simulated observation for this hypothetical scenario \citep{al-refaie.etal2021_TauRExIIIFast}. By treating the problem as one of optimization, the input parameters (planetary temperature, molecular abundances, cloud height, etc.) can be iterated so that a forward model of the simulated transit spectrum matches the observational data as closely as possible. However, this approach can be computationally intensive. Although the simplest of forward models are relatively lightweight, adding accurate cloud simulations increases physical accuracy at the cost of model complexity \citep{ma.etal2023_YunMaEnablingSpectral}, while including additional molecules increases comprehensiveness at the expense of the number of parameters to optimize \citep{welbanks.etal2025_ChallengesDetectionGases}. Both of these factors have a compounding impact on forward model inference speed. Furthermore, including additional molecules increases not only the complexity of the problem but also the risk of multiple degenerate solutions.

In order to minimize the number of calls to the forward model needed for convergence, the typical framework used is Bayesian sampling, such as MCMC \citep{foreman-mackey.etal2013_EmceeMCMCHammer, sharma2017_MarkovChainMonte} or nested sampling \citep{feroz.etal2009_MultiNestEfficientRobust, speagle2020_DynestyDynamicNested}. This concentrates samples in the regions of parameter space with high posterior probability (where the residuals between the observation and the simulation are small), and so maps the posterior with far fewer forward model calls than a naive grid search.

However, the field is shifting towards the use of nested sampling as the sampler instead, despite its worse convergence speed with large numbers of free parameters, as it has better performance across multiple degenerate solutions due to its ability to navigate multiple local minima concurrently, e.g. \textsc{TauREx-III} \citep{al-refaie.etal2021_TauRExIIIFast}. While MCMC explores the parameter space as a single agent in a random-walk fashion, nested sampling handles a set of live points concurrently, which effectively map the whole surface into a series of shells that can converge on multiple islands of solutions. On top of this, nested sampling provides \textit{Bayesian evidence}, which is a useful metric for native model comparison.

(Discussion continues in\Fig{sec:inference_from_exoplanet_obs}{Section}.)

\clearpage
\raggedright\setlength{\parindent}{0pt} 
\section{Details of retrievals for HD~209458~\texorpdfstring{\lowercase{b}}{b}}

\begin{tabular}{@{}lll@{}}
\textbf{Parameter }          & \textbf{Value }          & \textbf{Citation} \\
\hline\hline
Planet mass                  & 0.73 $M_\mathrm{J}$                    & \citep{stassun.etal2017_AccurateEmpiricalRadii}\\
Stellar radius                  & 1.19 $R_{\odot}$              & \citep{stassun.etal2017_AccurateEmpiricalRadii}\\
Stellar temperature             & 6091 K                        & \citep{stassun.etal2017_AccurateEmpiricalRadii}\\
Pressure bounds& $10^{-5}$ -- $10^6$ Pa& \\
He/$\mathrm{H_2}$ ratio               & 0.17                          & \\
CIA pairs                    & $\mathrm{H_2}$--$\mathrm{H_2}$, $\mathrm{H_2}$--He       & \\
\hline
Number of atmospheric layers& 100&\\
Number of live points& 1500&\\
\hline
\end{tabular}
\captionof{table}{Fixed parameters for retrieval of HD~209458~b}
\label{tab:hd209_params}

\begin{tabular}{lll}
\textbf{Retrieved Parameter}  & \textbf{Priors} & \textbf{Mode}   \\
\hline \hline
$\mathrm{H_2O}$               & $-12$, $-1$         & log    \\
$\mathrm{CH_4}$               & $-12$, $-1$         & log    \\
$\mathrm{CO}$                 & $-12$, $-1$         & log    \\
$\mathrm{CO_2}$               & $-12$, $-1$         & log    \\
$\mathrm{NH_3}$               & $-12$, $-1$         & log    \\
$T_\mathrm{isothermal}$ (K) & 400, 2000           & linear \\
Radius ($R_\mathrm{J}$)       & 1.2, 1.5            & linear \\
Cloud-top pressure (Pa) & -3, 6                & log   \\
\hline
\end{tabular}
\captionof{table}{Priors for retrieval of HD~209458~b}
\label{tab:hd209_priors}

{\renewcommand{\arraystretch}{1.3}%
\begin{tabular}{lll}
     \textbf{Retrieved Parameter}& \textbf{Median} & \textbf{Mean}\\
     \hline \hline
     $\log \mathrm{H_2O}$& $-3.63^{+1.18}_{-1.26}$& $-3.64$\\
     $\log \mathrm{NH_3}$& $-8.18^{+1.96}_{-2.39}$& $-8.27$\\
     $\log \mathrm{CO}$& $-7.48^{+3.20}_{-2.89}$& $-7.37$\\
     $\log \mathrm{CO_2}$& $-7.73^{+2.76}_{-2.67}$& $-7.67$\\
     $\log \mathrm{CH_4}$& $-8.12^{+2.64}_{-2.42}$& $-8.00$\\
    $T_\mathrm{isothermal}$ (K)& $838^{+386}_{-246}$& $896$\\
     $\log$ cloud-top pressure (Pa)& $2.96^{+1.59}_{-1.30}$& $3.08$\\
     Radius ($R_\mathrm{J}$)& $1.366^{+0.016}_{-0.026}$& $1.362$\\
     \hline
\end{tabular}}
\captionof{table}{Retrieved posteriors from the HD~209458~b retrieval, given as the posterior median with the 16th and 84th percentiles and as the posterior mean, both weighted by the nested-sampling weights.}
\label{tab:HD209_posteriors}

\clearpage
\section{Correlation matrix for HD~209458~\texorpdfstring{\lowercase{b}}{b}}

\includegraphics[width=0.48\textwidth]{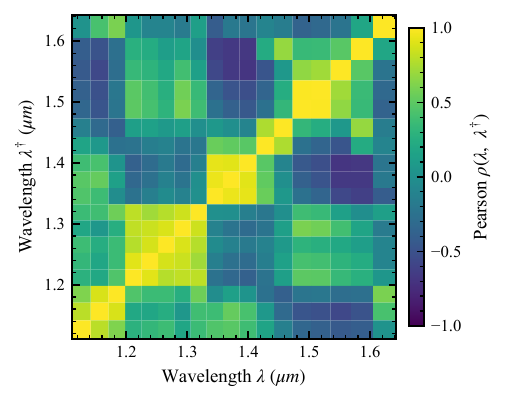}
\captionof{figure}{Pearson correlation matrix for all wavelength bins across the grid of forward models of HD~209458~b described in\fig{fig:Dm_logmol}. Correlation is calculated between pairs of wavelength bins in the min-max normalized spectra.}
\label{fig:correlation}

\onecolumn
\clearpage
\section{HD~209458~\texorpdfstring{\lowercase{b}}{b} metrics}
\label{sec:hd209_metrics}
\centering
\includegraphics[height=0.84\textheight]{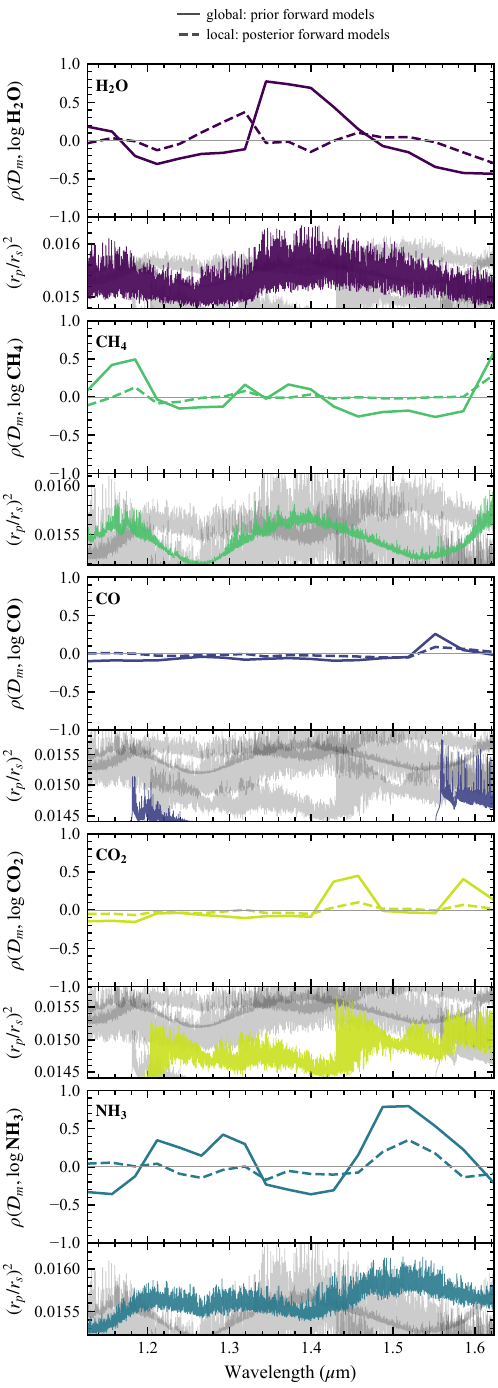}
\captionof{figure}{For each gas in the HD~209458~b retrieval, the Pearson correlation between the transit depth in each wavelength bin $m$ and the abundance of that gas across a set of forward models (upper panels), and the contribution of that gas to the spectrum (lower panels). The global metric uses 3\,125 forward models, with the abundances of the five gases on a regular grid from $10^{-10}$ to $10^{-2}$ in steps of 2\,dex, $T$, $R_\mathrm{p}$ and the cloud-top pressure fixed at 902.4\,K, 1.39\,$R_\mathrm{J}$ and $10^3$\,Pa, and each spectrum min-max normalized. The local metric uses forward models at the 8\,862 samples that hold 99.999 per cent of the weight of the \textsc{TauREx-III} posterior of the observation, on the raw spectra, weighted by the posterior. The contribution of a gas (in colour; the other gases in grey) is the spectrum of a forward model containing only that gas, at a volume mixing ratio of $10^{-2}$. The $\mathrm{H_2O}$ panels are shown in the main text in\fig{fig:Dm_logmol_H2O}.}
\label{fig:Dm_logmol}

\clearpage
\section{Full evaluation metrics comparison for HD~209458~\texorpdfstring{\lowercase{b}}{b}}
\centering
\includegraphics[width=0.86\textwidth]{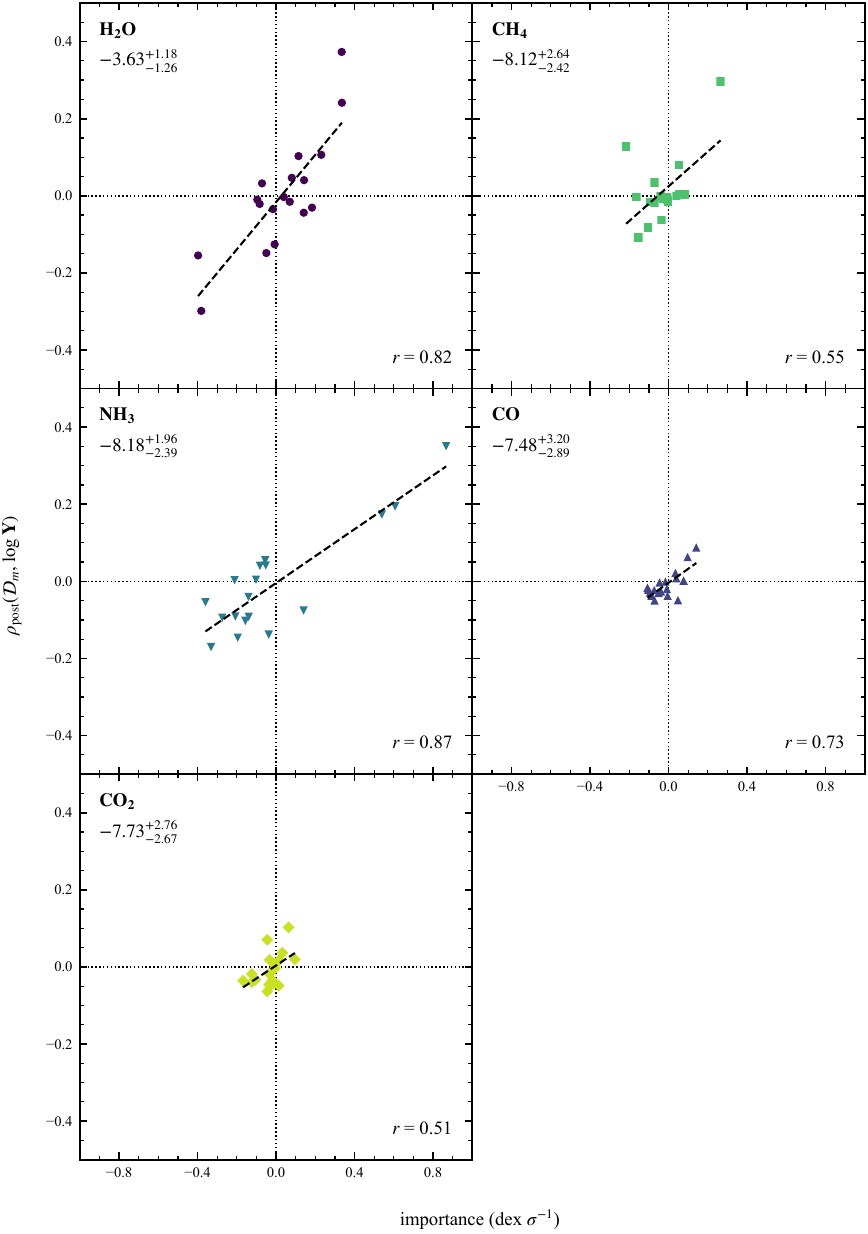}
\captionof{figure}{As\fig{fig:final_scatterplot}, for each gas in the retrieval: the importance of each wavelength bin against its local metric, the fitted linear trend and its Pearson $r$, and the retrieved $\log$ abundance (the posterior median with the 16th and 84th percentiles).}
\label{fig:final_scatterplot_all}

\twocolumn
\clearpage
\section{Fraction of the perturbation fitted for HD~209458~\texorpdfstring{\lowercase{b}}{b}}
\label{sec:fraction_fitted}

\includegraphics[width=0.48\textwidth]{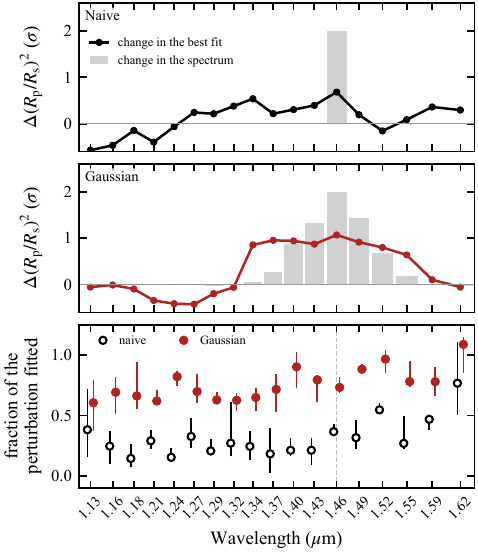}
\captionof{figure}{How much of each perturbation of the HD~209458~b spectrum the best-fitting \textsc{TauREx-III} model reproduces. The upper and middle panels show the change in the spectrum under naive and Gaussian perturbation of the $1.46\,\um$ bin by $+2\sigma$, and the change in the best-fitting model, both in units of each bin's uncertainty. The lower panel shows, for each perturbed bin, the fraction of the perturbation fitted\Figp{sec:hd209_naive}{Section}, as the median and interquartile range over the six perturbation magnitudes, for naive and Gaussian perturbation. The vertical dashed line marks the wavelength bin shown in the upper two panels.}
\label{fig:fraction_fitted}

\clearpage
\raggedright\setlength{\parindent}{0pt} 
\section{Details of retrievals for HAT-P-26~\texorpdfstring{\lowercase{b}}{b}}

\begin{tabular}{@{}lll@{}}
\textbf{Parameter }          & \textbf{Value }          & \textbf{Citation} \\
\hline\hline
Planet mass                  & 0.0585 $M_\mathrm{J}$ & \citep{gressier.etal2025_JWSTTSTDREAMSSulfur}\\
                             & (18.6 $M_{\oplus}$) & \\
Stellar radius                  & 0.87 $R_{\odot}$              & \citep{gressier.etal2025_JWSTTSTDREAMSSulfur}\\
Stellar temperature             & 5079 K                        & \citep{gressier.etal2025_JWSTTSTDREAMSSulfur}\\
Pressure bounds& $10^{-3}$ -- $10^{7}$ Pa& \\
He/$\mathrm{H_2}$ ratio               & 0.17                          & \\
CIA pairs                    & $\mathrm{H_2}$--$\mathrm{H_2}$, $\mathrm{H_2}$--He       & \\
\hline
Number of atmospheric layers& 100&\\
Number of live points& 1500&\\
\hline
\end{tabular}
\captionof{table}{Fixed parameters for retrieval of \mbox{HAT-P-26}~b}
\label{tab:hatp26_params}

\begin{tabular}{lll}
\textbf{Retrieved Parameter}  & \textbf{Priors} & \textbf{Mode}   \\
\hline \hline
$\mathrm{H_2O}$               & $-12$, $-0.3$       & log    \\
$\mathrm{CH_4}$               & $-12$, $-0.3$       & log    \\
$\mathrm{CO}$                 & $-12$, $-0.3$       & log    \\
$\mathrm{CO_2}$               & $-12$, $-0.3$       & log    \\
$\mathrm{NH_3}$               & $-12$, $-0.3$       & log    \\
$\mathrm{H_2S}$               & $-12$, $-0.3$       & log    \\
$\mathrm{HCN}$                & $-12$, $-0.3$       & log    \\
$\mathrm{C_2H_2}$             & $-12$, $-0.3$       & log    \\
$\mathrm{C_2H_4}$             & $-12$, $-0.3$       & log    \\
$\mathrm{SO_2}$                & $-12$, $-0.3$       & log    \\
$T_\mathrm{isothermal}$ (K) & 500, 1500           & linear \\
Radius ($R_\mathrm{J}$)       & 0.3, 0.9            & linear \\
Cloud-top pressure (Pa)  & $-3$, 7& log   \\
\hline
\end{tabular}
\captionof{table}{Priors for retrieval of \mbox{HAT-P-26}~b}
\label{tab:hatp26_priors}

{\renewcommand{\arraystretch}{1.3}%
\begin{tabular}{lll}
     \textbf{Retrieved Parameter}& \textbf{Median} & \textbf{Mean}\\
     \hline \hline
     $\log \mathrm{H_2O}$& $-1.51^{+0.20}_{-0.34}$& $-1.63$\\
     $\log \mathrm{CO_2}$& $-2.68^{+0.41}_{-0.49}$& $-2.77$\\
     $\log \mathrm{SO_2}$& $-4.72^{+0.44}_{-0.50}$& $-4.78$\\
     $\log \mathrm{NH_3}$& $-8.96^{+2.05}_{-1.93}$& $-8.89$\\
     $\log \mathrm{CO}$& $-7.83^{+2.84}_{-2.67}$& $-7.73$\\
     $\log \mathrm{CH_4}$& $-8.23^{+2.00}_{-2.44}$& $-8.38$\\
     $\log \mathrm{C_2H_2}$& $-9.39^{+1.81}_{-1.65}$& $-9.31$\\
     $\log \mathrm{C_2H_4}$& $-8.62^{+1.94}_{-2.14}$& $-8.70$\\
     $\log \mathrm{H_2S}$& $-3.65^{+0.60}_{-4.36}$& $-4.91$\\
     $\log \mathrm{HCN}$& $-8.99^{+1.95}_{-1.91}$& $-8.94$\\
    $T_\mathrm{isothermal}$ (K)& $600^{+68}_{-56}$& $605$\\
     $\log$ cloud-top pressure (Pa)& $3.43^{+2.32}_{-1.14}$& $3.85$\\
     Radius ($R_\mathrm{J}$)& $0.5444^{+0.0075}_{-0.0107}$& $0.5431$\\
     \hline
\end{tabular}}
\captionof{table}{Retrieved posteriors from the \mbox{HAT-P-26}~b retrieval, as in\Fig{tab:HD209_posteriors}{Table}. The temperature, radius and cloud-top pressure of the training set of\Fig{sec:ml}{Section}, and the abundances of the simulated observation it is tested on, are the posterior means.}
\label{tab:HATP_posteriors}

\clearpage
\onecolumn
\section{HAT-P-26~\texorpdfstring{\lowercase{b}}{b} metrics}

\centering
\includegraphics[height=0.84\textheight]{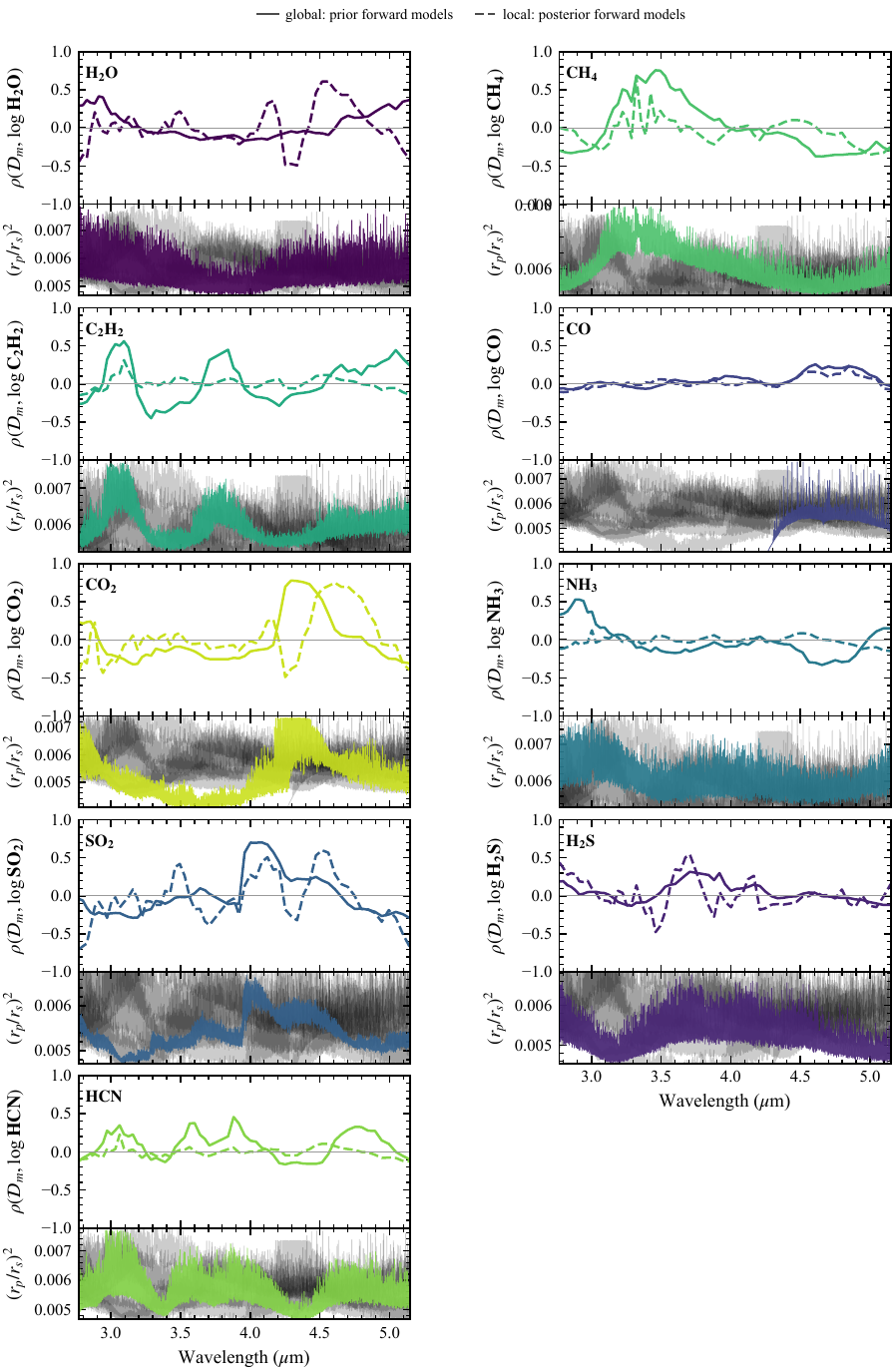}
\captionof{figure}{Repeat of\fig{fig:Dm_logmol} for \mbox{HAT-P-26}~b. The global metric uses 486 forward models in which the abundances of nine of the ten gases (all but $\mathrm{C_2H_4}$) are drawn from the retrieval priors, with $T$, $R_\mathrm{p}$ and the cloud-top pressure fixed at 701\,K, 0.54\,$R_\mathrm{J}$ and $10^{2.55}$\,Pa. The local metric uses forward models at 4\,000 draws from the \textsc{TauREx-III} posterior of the observation.}
\label{fig:HATP_correlation_contribution}

\clearpage
\section{Full posteriors}
\label{sec:full_posteriors}
\centering
\includegraphics[width=0.98\textwidth]{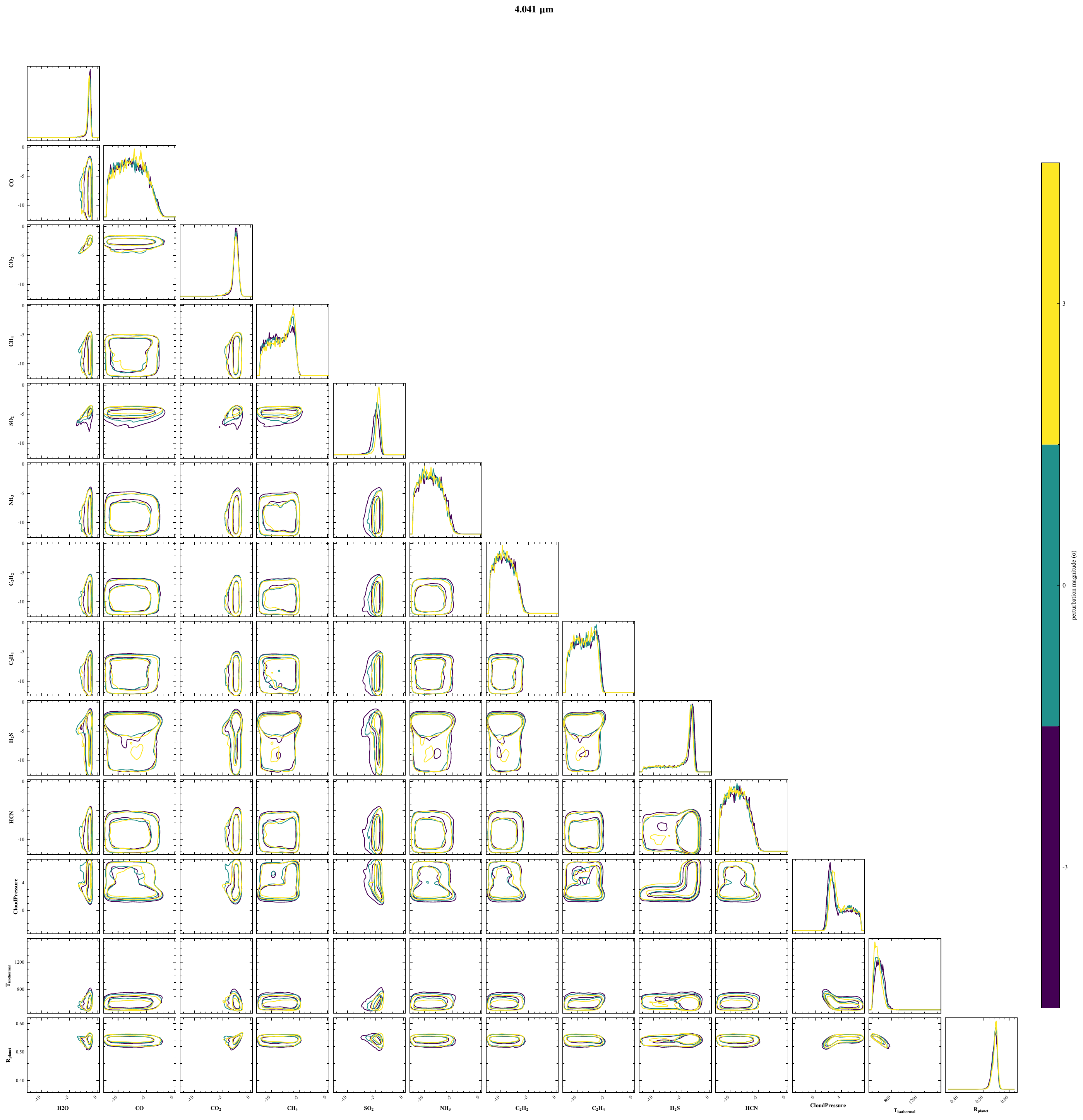}
\captionof{figure}{Full posterior of the \textsc{TauREx-III} retrieval of \mbox{HAT-P-26}~b under naive perturbation of the $4.04\,\um$ bin alone by $-3\sigma$ (purple), $0\sigma$ (teal, the unperturbed retrieval) and $+3\sigma$ (yellow). The $\mathrm{SO_2}$ posterior shifts in the direction of the perturbation, while those of $\mathrm{CO_2}$ and $\mathrm{H_2O}$ barely move.}
\label{fig:4.041_HATP_full_posterior_naive}

\clearpage
\centering
\includegraphics[width=0.98\textwidth]{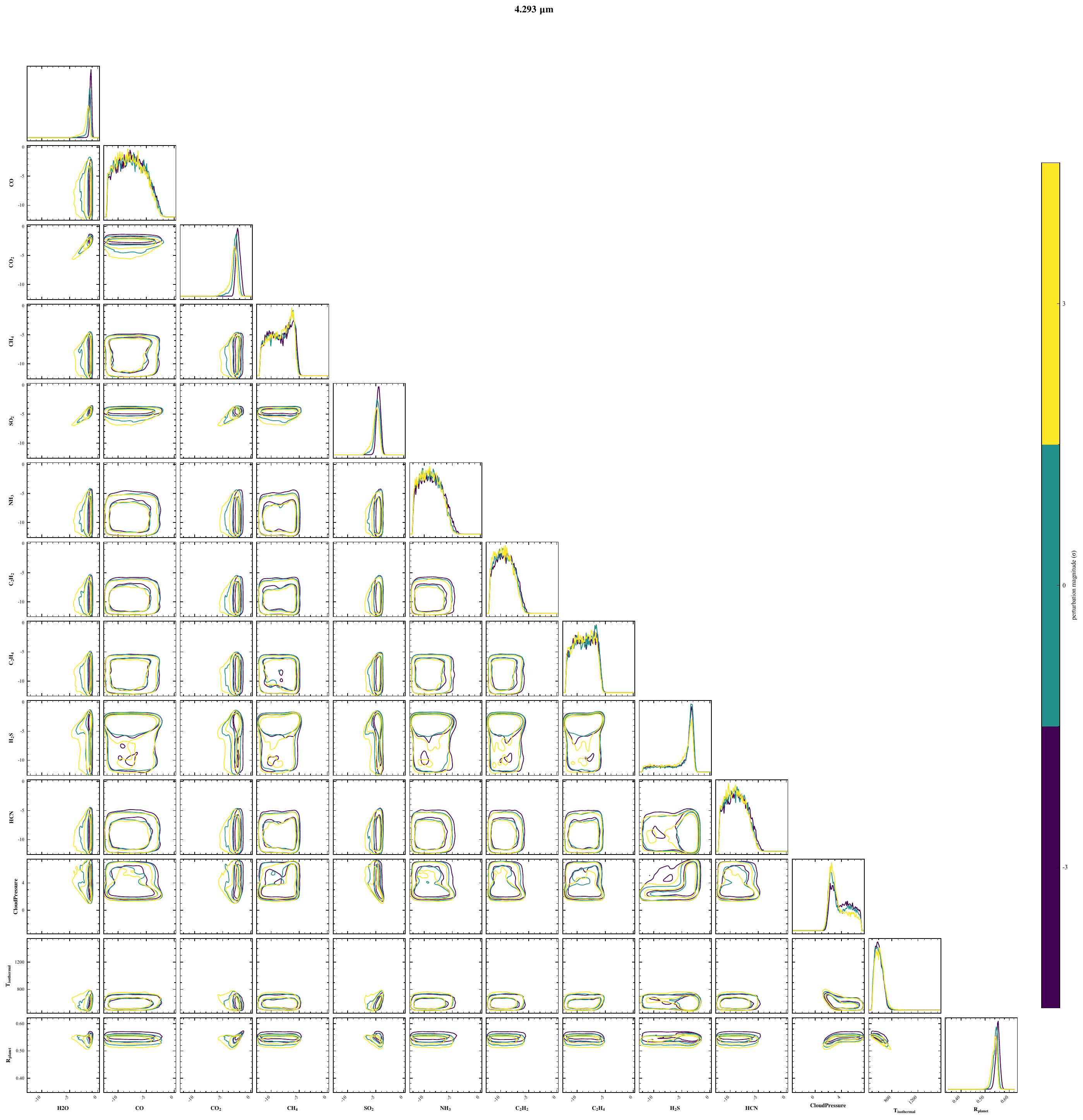}
\captionof{figure}{Full posterior of the \textsc{TauREx-III} retrieval of \mbox{HAT-P-26}~b under naive perturbation of the $4.29\,\um$ bin alone by $-3\sigma$ (purple), $0\sigma$ (teal, the unperturbed retrieval) and $+3\sigma$ (yellow). As under the Gaussian perturbation\figp{fig:4.293_HATP_full_posterior_gauss}, but by less, the $\mathrm{CO_2}$, $\mathrm{SO_2}$ and $\mathrm{H_2O}$ posteriors all shift against the direction of the perturbation.}
\label{fig:4.293_HATP_full_posterior_naive}

\clearpage
\centering
\includegraphics[width=0.98\textwidth]{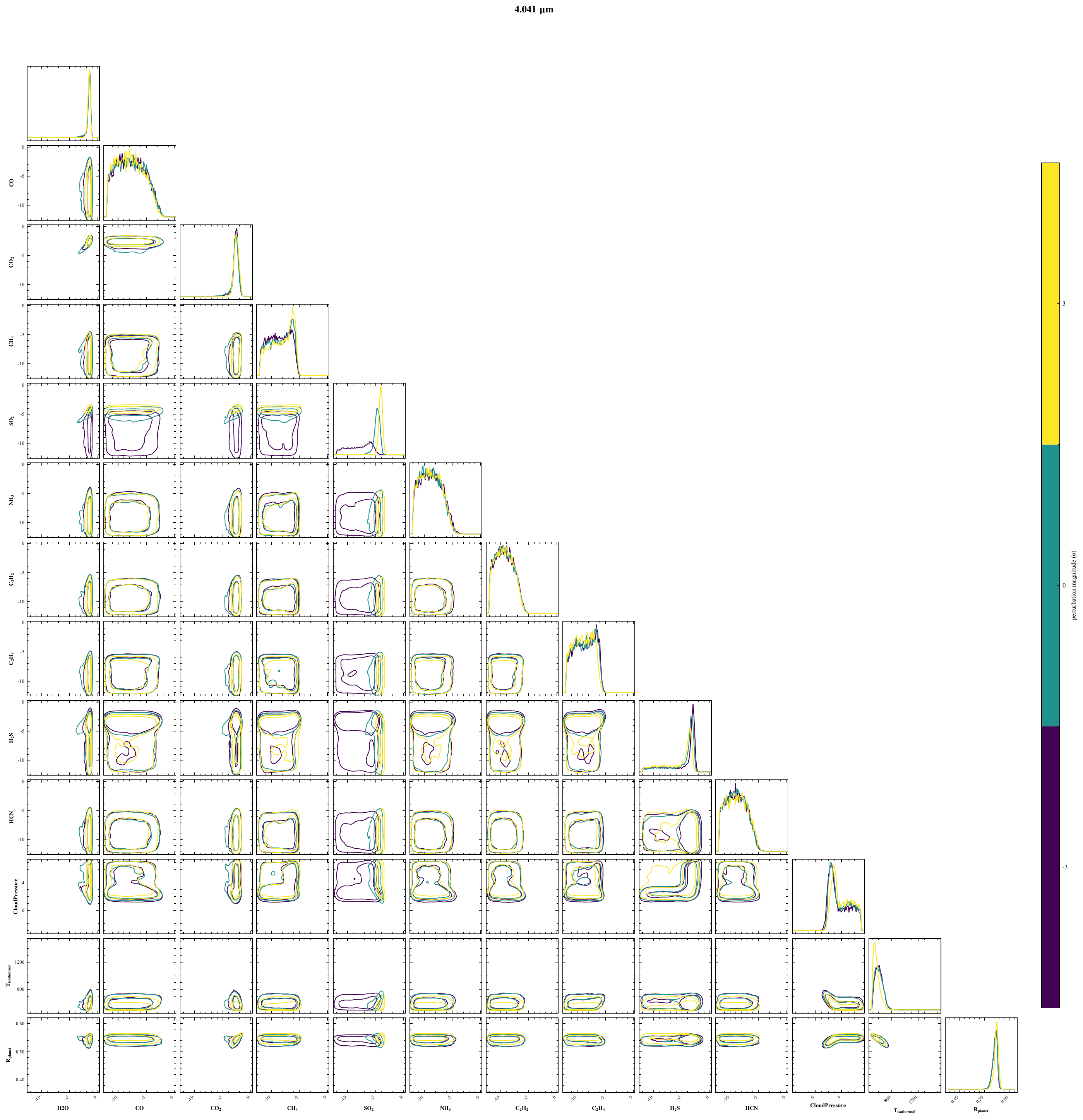}
\captionof{figure}{Full posterior of the \textsc{TauREx-III} retrieval of \mbox{HAT-P-26}~b under Gaussian perturbation of $-3\sigma$ (purple), $0\sigma$ (teal, the unperturbed retrieval) and $+3\sigma$ (yellow) centred on $4.04\,\um$. The $\mathrm{SO_2}$ posterior shifts in the direction of the perturbation, and under $-3\sigma$ it extends to much lower abundances, while those of $\mathrm{CO_2}$ and $\mathrm{H_2O}$ barely move.}
\label{fig:4.041_HATP_full_posterior_gauss}

\clearpage
\centering
\includegraphics[width=0.98\textwidth]{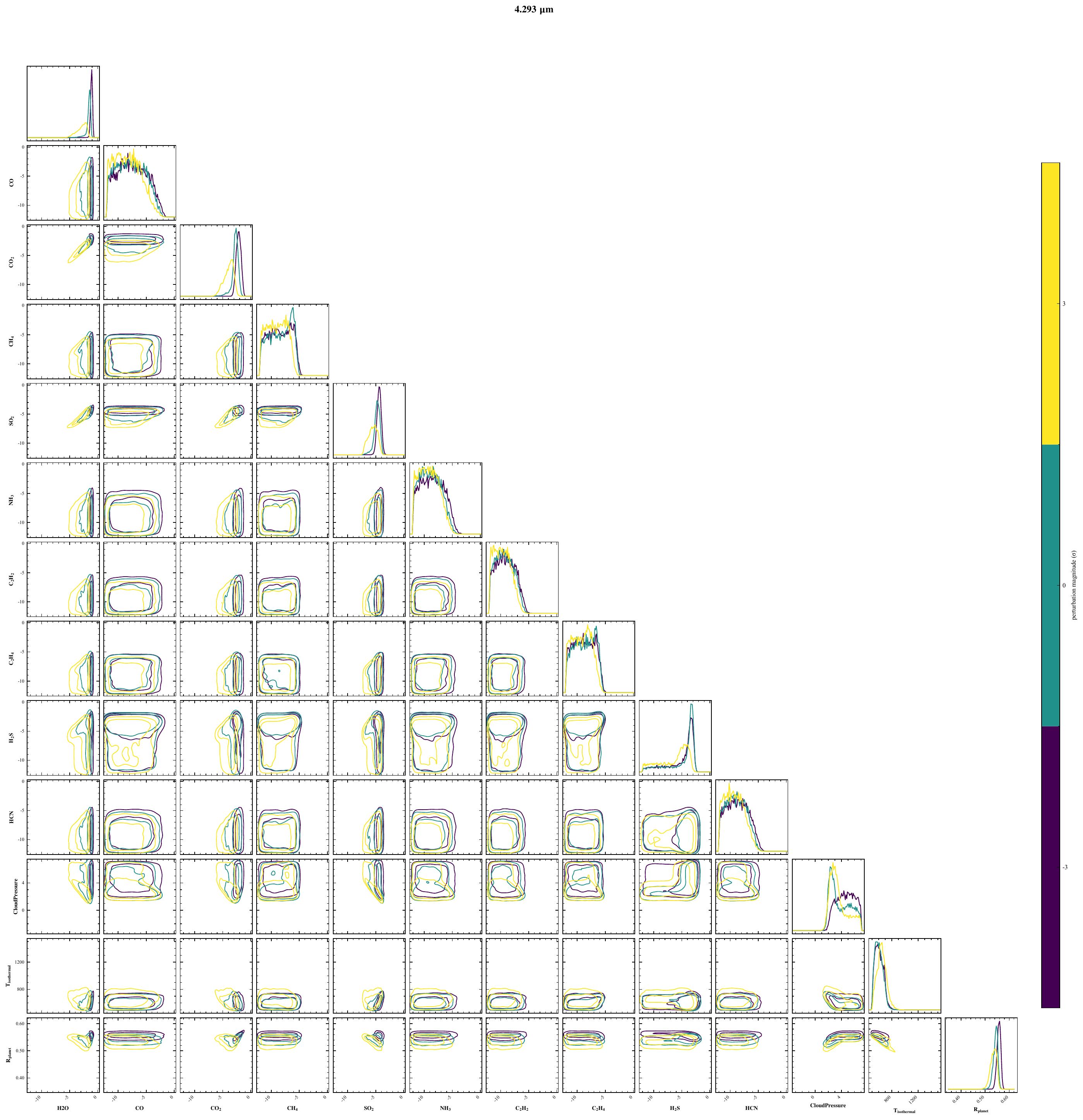}
\captionof{figure}{Full posterior of the \textsc{TauREx-III} retrieval of \mbox{HAT-P-26}~b under Gaussian perturbation of $-3\sigma$ (purple), $0\sigma$ (teal, the unperturbed retrieval) and $+3\sigma$ (yellow) centred on $4.29\,\um$. Under $+3\sigma$ the posterior extends to lower $\mathrm{CO_2}$ and $\mathrm{H_2O}$ and higher $T$.}
\label{fig:4.293_HATP_full_posterior_gauss}

\twocolumn
\clearpage
\raggedright
\section{Details of the machine learning model used in retrievals for HAT-P-26~\texorpdfstring{\lowercase{b}}{b}}
\label{sec:cnn}

\subsection{Network architecture}
 The simple one-dimensional CNN used maps a transmission spectrum directly to the $\log$ volume mixing ratios (VMRs) of the ten free gases. Its architecture is that of the Monte Carlo Dropout baseline released for the ADC 2022 \citep{yip.etal2022_ESAArielDataChallenge},\footnote{\url{https://github.com/ucl-exoplanets/NeurIPS2022_Baseline}} itself adapted from the CNN of \citet{yipPeekingBlackBox2021a}. The 60-bin spectrum is treated as a single-channel sequence and passed through three convolutional blocks, each of two convolutional layers with a three-bin kernel and rectified linear unit (ReLU) activations followed by max pooling over two bins. The number of filters doubles from one block to the next (32, 64, 128) while the sequence is condensed from 60 to 4 positions, each of which draws on 36 adjacent bins in the original input spectrum. These 512 hidden features are flattened and passed through two fully connected layers of 500 and 100 units, each followed by dropout, and a linear output layer that returns the ten normalized log abundances (403\,386 trainable parameters in total).

\subsection{Training}
The training data are drawn from 100\,000 transmission spectra computed with \textsc{TauREx-III} and binned on to the 60 wavelength bins of the \mbox{HAT-P-26}~b NIRSpec G395H spectrum. Because we only make predictions for this one planet, every spectrum uses the same stellar parameters and planetary mass\Figp{tab:hatp26_params}{Table}, and the same isothermal temperature (605\,K), planetary radius (0.5431\,$R_\mathrm{J}$) and grey cloud-top pressure ($10^{3.85}$\,Pa) from the posterior means retrieved by \textsc{TauREx-III}\Figp{tab:HATP_posteriors}{Table}. The network therefore needs none of the auxiliary inputs of the ADC baseline (the stellar and planetary radii), since these parameters are constant across the training set. The log abundances of $\mathrm{H_2O}$, $\mathrm{CO_2}$, $\mathrm{SO_2}$, $\mathrm{H_2S}$, $\mathrm{CO}$, $\mathrm{CH_4}$, $\mathrm{NH_3}$, $\mathrm{HCN}$, $\mathrm{C_2H_2}$ and $\mathrm{C_2H_4}$ are drawn independently from the retrieval prior, $\mathcal{U}(-12, -0.3)$\Figp{tab:hatp26_priors}{Table}, rejecting draws whose trace gases sum to a VMR above 0.85 so that the $\mathrm{H_2/He}$ fill stays positive. The spectra are split at random into training, validation and test sets in the ratio 8:1:1 respectively.

Each wavelength bin of the input, and each gas of the output, is mapped linearly so that the 0.5th and 99.5th percentiles of the training set fall at 0 and 1. These maps are not clipped. The stored spectra are noiseless, and noise is added only when the network is trained or evaluated. Similarly to the ADC baseline, which trains on 10 fixed noisy copies of each spectrum, we add Gaussian noise with standard deviation $\sigma_m$, the uncertainty of the observed spectrum in bin $m$, to every training spectrum. However, we redraw this noise at every epoch in order to further increase the robustness of the network. The validation loss is assessed on five fixed noisy realizations of each validation spectrum. Training (and evaluating) a model with noisy data is critical to demonstrate our interpretability methods, as the whole principle of applying perturbations of order $\sigma_m$ to remain within the training distribution is no longer applicable if $\sigma_m = 0$ across a noiseless training distribution. Gaussian perturbation is only \textit{more effective than} naive perturbation at making samples that look in-distribution: there is still a level of deviation tolerance required from the model, which is absent in a noiseless training scenario.

The network is implemented in \textsc{PyTorch} \citep{paszke.etal2019_PyTorchImperativeStyle} and trained with the Adam optimizer \citep{kingma.ba2015_AdamMethodStochastic} to minimize the mean squared error (MSE) on the normalized targets, keeping the baseline's loss, initial learning rate ($10^{-3}$) and dropout rate (0.1);\Fig{tab:cnn_hyperparameters}{Table} lists the full set of hyperparameters used, alongside the comparable values from the ADC baseline. We increase the batch size from 32 to 128 due to the larger training set. Rather than limiting the number of training epochs, we implement early stopping and a dynamic learning rate: the learning rate is halved after 10 epochs without improvement, and training stops after 30. Training stopped at epoch 276 (15 minutes on a \texttt{Mac M4 Pro 48\,GB} using \textsc{mps}), and we keep the weights from epoch 246. As expected, accuracy in the trained network varies between gases, from $R^2 = 0.65$ for $\mathrm{C_2H_4}$ and 0.56 for $\mathrm{CO_2}$ to 0.15 for $\mathrm{CO}$, and errors are largest for molecules whose features have a low signal-to-noise ratio, such as $\mathrm{H_2S}$. These scores are the same, to the precision quoted, whether dropout is switched off at inference or averaged over 1\,000 masks.

\begin{table}
    \centering
 \setlength{\tabcolsep}{4pt}
  \begin{tabular}{lll}
    \hline
    Hyperparameter & This work & ADC baseline \\
    \hline
    \multicolumn{3}{l}{\textit{Architecture}} \\
    Inputs & 60 bins & 52 bins, $R_\mathrm{s}$, $R_\mathrm{p}$ \\
    Outputs & 10 log-VMRs & $T$, 5 log-VMRs \\
    Convolutional blocks & \multicolumn{2}{c}{3, each 2 conv.\ layers + max pool} \\
    Filters per block & \multicolumn{2}{c}{32, 64, 128} \\
    Kernel size, stride & \multicolumn{2}{c}{3, 1 (no padding)} \\
    Pool size & \multicolumn{2}{c}{2} \\
    Dense layers & \multicolumn{2}{c}{500, 100 units} \\
    Activation & \multicolumn{2}{c}{ReLU; linear output} \\
    Dropout rate & \multicolumn{2}{c}{0.1, after each hidden dense layer} \\
    Dropout at inference & \multicolumn{2}{c}{Monte Carlo dropout}\\
    \hline
    \multicolumn{3}{l}{\textit{Data}} \\
    Training spectra & 80\,015 & $\sim$4\,000 \\
    Validation spectra & 9\,991 & $\sim$1\,000 \\
    Input scaling & per-bin percentile$^1$ & global z-score \\
    Target scaling & per-gas percentile$^1$ & per-target z-score \\
    Training noise & $\sigma_m$, redrawn each epoch & $\sigma_m$, 10 fixed copies \\
    Validation noise & $\sigma_m$, 5 fixed copies & none \\
    \hline
    \multicolumn{3}{l}{\textit{Optimization}} \\
    Loss & \multicolumn{2}{c}{MSE} \\
    Optimizer & \multicolumn{2}{c}{Adam ($\beta_1 = 0.9$, $\beta_2 = 0.999$)} \\
    Initial learning rate & \multicolumn{2}{c}{$10^{-3}$} \\
    Weight decay & \multicolumn{2}{c}{0} \\
    Batch size & 128 & 32 \\
    Shuffling & each epoch & none \\
    Maximum epochs & None & 10 \\
    Learning-rate schedule & halved on plateau$^2$ & constant \\
    Early stopping & 30-epoch patience & none \\
    Retained weights & best validation loss & final epoch \\
    Epochs run (retained) & 276 (246) & 10 (10) \\
    \hline
  \end{tabular}
    \caption{Hyperparameters of the CNN retrieval\Figp{sec:ml}{Section} and of the ADC 2022 Monte Carlo Dropout baseline from which it is derived \citep{yip.etal2022_ESAArielDataChallenge}. Entries spanning both columns are shared. Noise is Gaussian with standard deviation $\sigma_m$, the observational uncertainty in wavelength bin $m$. $^1$Linear map of the training-set 0.5th and 99.5th percentiles to 0 and 1. $^2$Halved after 10 epochs without improvement in the validation loss. }
    \label{tab:cnn_hyperparameters}
\end{table}

\onecolumn
\clearpage
\centering

\section{Full posterior for the machine learning retrieval of HAT-P-26~\texorpdfstring{\lowercase{b}}{b}}

\includegraphics[width=0.98\textwidth]{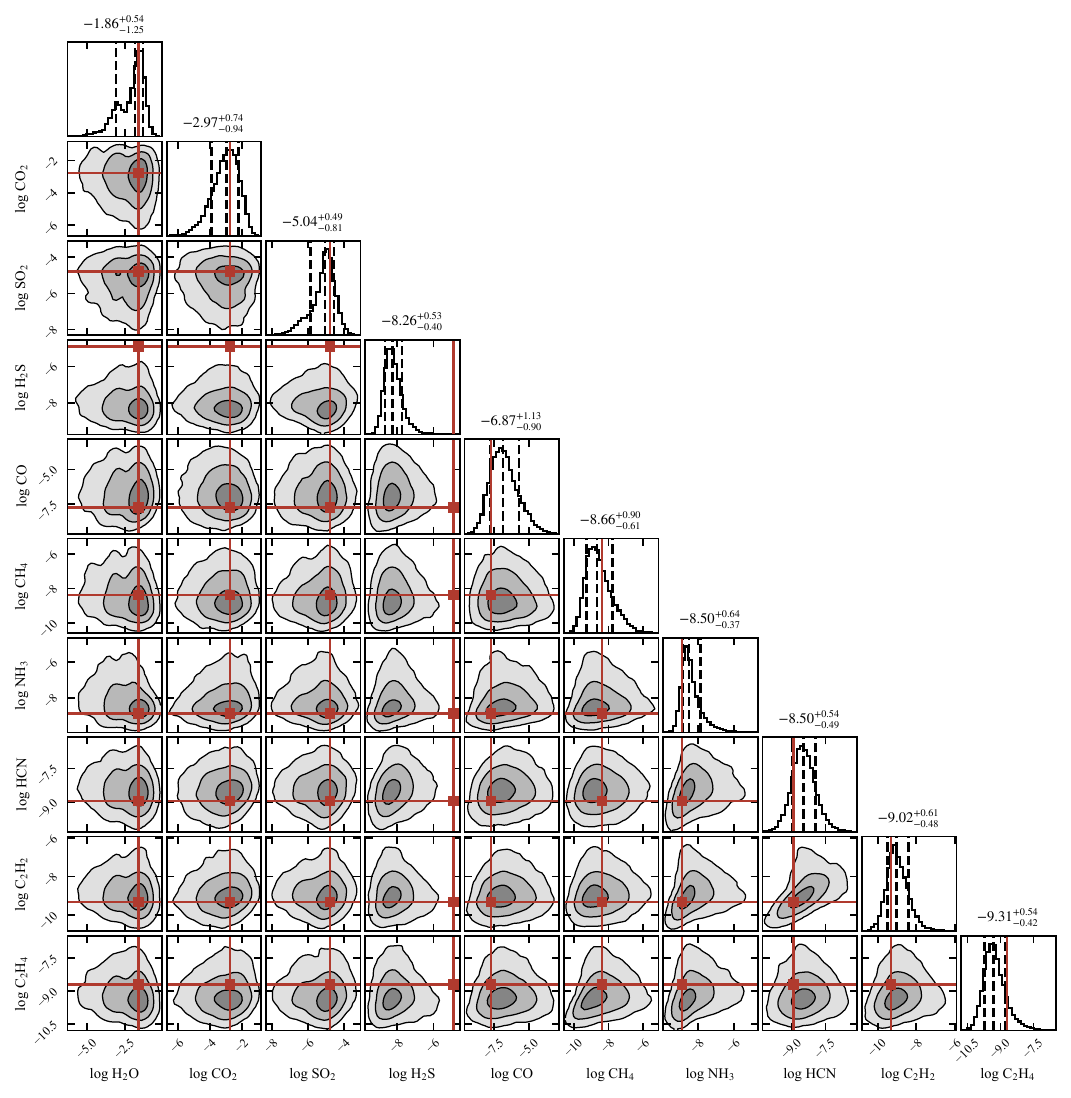}
\captionof{figure}{Full posterior of the CNN retrieval of the simulated \mbox{HAT-P-26}~b spectrum. 20\,000 noisy realizations of the spectrum are each passed through the network once with an independent Monte Carlo dropout mask. Contours enclose 1, 2 and 3$\sigma$; dashed lines mark the median and 1$\sigma$ range, quoted above each column; red lines mark the input values (the \textsc{TauREx-III} posterior means of\Fig{tab:HATP_posteriors}{Table}).}
\label{fig:HATP_full_posterior_ML}

\clearpage
\section{Fixed and fresh Monte Carlo dropout masks}
\label{sec:fresh_dropout}
\centering
\includegraphics[width=0.98\textwidth]{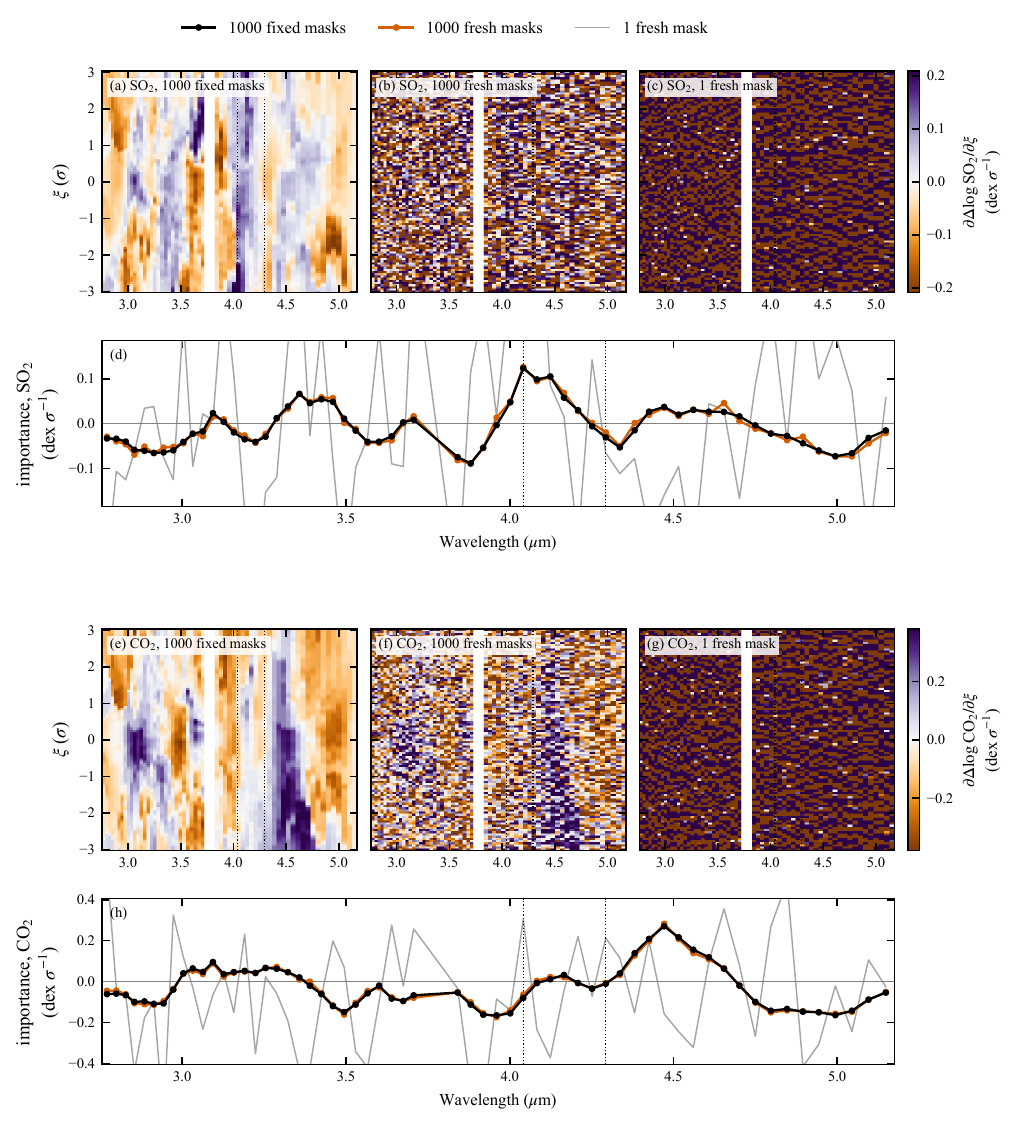}
\captionof{figure}{Response-gradient maps and importance of the CNN retrieval of the simulated \mbox{HAT-P-26}~b spectrum, for $\mathrm{SO_2}$ (a--d) and $\mathrm{CO_2}$ (e--h), with fixed and with fresh Monte Carlo dropout masks. (a, e) The maps of\fig{fig:hatp_corner_co2_so2}, in which every evaluation of $f$ is the mean over the same 1\,000 dropout masks. (b, f) Every evaluation is the mean over 1\,000 fresh masks. (c, g) Every evaluation is a single pass with a fresh mask. Colour scales are shared within each molecule and saturate at the 99th percentile of the fixed-mask map. (d, h) Importance, the mean response gradient over the grid, in each case; the single-mask curve extends beyond the axes. With 1\,000 fresh masks the importance matches the fixed-mask importance in every bin, but the response-gradient maps are dominated by noise. Dotted lines mark the $\mathrm{SO_2}$ and $\mathrm{CO_2}$ features at $4.04\,\um$ and $4.29\,\um$.}
\label{fig:fresh_dropout}


\end{document}